\documentclass[journal]{IEEEtran}
\usepackage{cite}
\usepackage[compatible]{algpseudocode}
\usepackage{amsmath,amsfonts, amssymb, amsthm}
\usepackage{graphicx}
\usepackage{mathtools}
\usepackage[ruled,vlined, linesnumbered]{algorithm2e}
\usepackage{enumerate}
\usepackage[shortlabels]{enumitem}

\newtheorem{theorem}{Theorem}

\usepackage{array}
\usepackage[table,xcdraw]{xcolor}
\usepackage{textcomp}
\usepackage{stfloats}
\usepackage{subcaption}
\usepackage{enumitem}
\usepackage{url}
\usepackage[hidelinks]{hyperref}
\usepackage{verbatim}
\usepackage{tikz}

\def\BibTeX{{\rm B\kern-.05em{\sc i\kern-.025em b}\kern-.08em
    T\kern-.1667em\lower.7ex\hbox{E}\kern-.125emX}}
\usepackage{balance}

\newtheorem*{assumption*}{\assumptionnumber}
\providecommand{\assumptionnumber}{}
\makeatletter
\newenvironment{assumption}[2]
 {%
  \renewcommand{\assumptionnumber}{Assumption #1}%
  \begin{assumption*}%
  \protected@edef\@currentlabel{#1-$\mathcal{#2}$}%
 }
 {%
  \end{assumption*}
 }
\makeatother

\begin{document}
\title{FedSem: A Resource Allocation Scheme for Federated Learning Assisted Semantic Communication}

\author{Xinyu~Zhou, Yang~Li, Jun~Zhao
\thanks{Xinyu Zhou and Yang Li are with ERI@N, Interdisciplinary Graduate Programme, Nanyang Technological University, Singapore. All authors are with the College of Computing and Data Science (CCDS), Nanyang Technological University (NTU), Singapore.  Emails: xinyu003@e.ntu.edu.sg, yang048@e.ntu.edu.sg, JunZhao@ntu.edu.sg
} 
}

\markboth{}%
{}

\IEEEtitleabstractindextext{%
\begin{abstract}
Semantic communication (SemCom), regarded as the evolution of the traditional Shannon's communication model, stresses the transmission of semantic information instead of the data itself.
Federated learning (FL), owing to its distributed learning and privacy-preserving properties, has received attention from both academia and industry.
In this paper, we introduce a system that integrates FL and SemCom, which is called FedSem. We have also proposed an optimization problem related to resource allocation for this system. The objective of this problem is to minimize the energy consumption and delay of FL, as well as the transmission energy of SemCom, while maximizing the accuracy of the model trained through FL.
The channel access scheme is Orthogonal Frequency-Division Multiple Access (OFDMA).
The optimization variables include the binary (0-1) subcarrier allocation indicator, the transmission power of each device on specific subcarriers, the computational frequency of each participating device, and the compression rate for SemCom.
To tackle this complex problem, we propose a resource allocation algorithm that decomposes the original problem into more tractable subproblems. By employing convex optimization techniques, we transform the \mbox{non-convex} problem into convex forms, ensuring tractability and solution effectiveness. Our approach includes a detailed analysis of time complexity and convergence, proving the practicality of the algorithm.
Numerical experiments validate the effectiveness of our approach, showing superior performance of our algorithm in various scenarios compared to baseline methods. Hence, our solution is useful for enhancing the operational efficiency of FedSem systems, offering significant potential for real-world applications. \vspace{2pt}
\end{abstract}

\begin{IEEEkeywords}
Resource allocation, federated learning, wireless communication, semantic communication, OFDMA, \mbox{non-convex} optimization. \vspace{-5pt}
\end{IEEEkeywords}}

\maketitle

\IEEEdisplaynontitleabstractindextext

% \fi
\IEEEpeerreviewmaketitle

\section{Introduction}

\subsection{Background and Motivation\vspace{1pt}}
In this age where mobile devices and communication scenarios are exploding, communication technology constantly evolves. The advent of the fifth generation (5G) has put more emphasis on low latency and high transmission rates. However, with the emergence of augmented reality (AR), virtual reality (VR), and the notion of Metaverse, the existing communication technologies are far from meeting the requirements of these scenarios. Hence, many scholars have looked beyond the traditional Shannon's communication model to investigate other possible communication architectures, such as semantic communication (SemCom) \cite{shi2021fromsemcom}.

SemCom has been deemed a revolutionary development of the traditional communication model. 
% Traditional communication focuses primarily on the technical aspects of transmitting information –- quantifying the maximum amount of data that can be reliably transmitted over a given communication channel.
Traditional communication focuses on the technical aspects of data transmission, particularly maximizing reliable data transfer over a given channel.
In contrast, SemCom provides a different perspective emphasizing the transmission of meaning and context.
This approach considers not just the efficient transmission of data but also how the information is interpreted and understood by the receiver.
Usually, in SemCom, there will be a semantic encoder that encodes the transmitted information to extract the semantic features of the original input data. 
Then, a channel encoder will be used to compress the information for robustness against noise in the physical channel.
Then, the compressed information will be delivered through the physical channel. 
A channel decoder on the receiver will first extract the semantic features, and a semantic decoder will recover the data as accurately as possible \cite{luo_sem_2022}. 
% This approach is increasingly promising today as we move towards more nuanced and intelligent communication systems capable of understanding and interpreting human language and intentions. 
This approach is increasingly promising as communication systems become more intelligent and able to understand and interpret human language and intentions.

% However, with the increasing concern for user privacy and data security (e.g., the General Data Protection Regulation passed by the European Union), training a SemCom model by collecting all users' data has become more challenging than ever before.
However, growing concerns about privacy and data security (e.g., the EU's General Data Protection Regulation) make training a SemCom model with collected user data more challenging than ever.
Hence, we consider incorporating federated learning (FL) into the SemCom system because it offers a distributed approach to training SemCom models without the need for data sharing.
FL, proposed by Google \cite{mcmahan2017communication}, is a distributed machine learning paradigm where models are trained collaboratively across multiple devices without sharing raw data.

Specifically, the integration of FL and SemCom in this paper provides several key benefits:
\begin{itemize}
    \item \textbf{Privacy Protection:} Traditional SemCom training relies on centralized data aggregation, increasing the risk of data breaches and compliance issues. FL could mitigate the above issues by enabling on-device training and provide privacy-preserving training while aligning with GDPR and other regulations.
    \item \textbf{Improved Generalization:}  Training a SemCom model on a centralized
    data source may limit its ability to generalize across different data distributions. In this case, FL allows to leverage of decentralized user data, which could enhance the global model’s adaptability across different users and applications.
    \item \textbf{Scalability \& Efficiency} As SemCom applications scale, centralized training may become inefficient. In this case, FL distributes training across devices, reducing server load and making large-scale deployment more feasible and resource-efficient, especially in wireless networks and edge computing scenarios.
\end{itemize}

% Additionally, FL facilitates efficient use of network resources and scalable model training across numerous devices, making it suitable for large-scale and diverse communication environments.
FL optimizes network resources and enables scalable model training, making it ideal for large and diverse communication environments.
Also, considering that data is dynamic and constantly changing rather than static, it is necessary to regularly update the model. 
FL enhances convenience in this process. 
Thus, the integration of FL and SemCom can help the model update more efficiently. 
FL process can utilize the local computation resources for training and updating the model.
Thus, the combination of FL and SemCom does bring several benefits. Nevertheless, as the number of mobile devices is growing in contemporary society, the generated data is proliferating, which results in challenges related to bandwidth, latency, energy consumption, etc. This paper mainly focuses on the challenges regarding wireless communications.

% In conclusion, the combination of FL with SemCom has the following benefits:

% FL enables training models on local data without centralizing sensitive information, which is crucial for maintaining privacy and security.

% FL facilitates efficient use of network resources and scalable model training across numerous devices, making it suitable for large-scale and diverse communication environments. 

% FL could enable context-aware, personalized models through decentralized learning from diverse local data, improving linguistic and cultural understanding and adaptability.

% 1) 

% % It allows for the development of distributed SemCom models that are trained on diverse, \mbox{real-world} data without compromising user privacy. 
% 2) 
% 3) 

\subsection{Challenges}
In this subsection, the detailed challenges related to wireless communications are discussed.

First, the energy consumption is a significant concern. Mobile devices such as smartphones and wearable devices have limited battery life. FL needs to perform local computations for model training and transmit updates, and SemCom requires suitable models to understand and transmit meaningful information, which all demand substantial energy consumption. The combination will potentially escalate the energy. Second, time efficiency is another concern. For FL, the latency in training and updating models can be significant, especially if the devices have various computational capabilities. SemCom also needs time to encode and decode semantic information, which may be substantial. 
% Third, since optimizing the energy and time consumption will affect the model training, how to balance the model accuracy and the energy and time consumption is challenging.

Third, since optimizing energy and time consumption impacts SemCom model training, balancing model accuracy with energy and time consumption presents a challenge. 
In the context of SemCom, model accuracy depends on selecting an appropriate compression rate for semantic information transmission. 
The compression rate in this paper refers to the ratio of transmitted information to the original information.
A lower compression rate reduces energy and transmission time but may degrade the accuracy of the transmitted information. 
Similarly, a higher compression rate preserves accuracy but increases energy consumption and latency. Additionally, in FL, local training consumes computation resources, while communication overhead adds further constraints. The interplay between these factors introduces a complex optimization problem, as an aggressive reduction in energy or time may affect overall system performance. Addressing this challenge requires an effective resource allocation strategy that dynamically adjusts computational and communication resources to achieve an optimal trade-off among accuracy, energy and latency.
Besides, we call the system that combines FL and SemCom, FedSem. The channel access scheme we use is orthogonal frequency-division multiple access (OFDMA).

The optimization problem is to minimize the total energy and time consumption and maximize the model accuracy of the whole FedSem system. Moreover, the optimization variables include the transmission power, allocated subcarriers and compression rate of each device. The compression rate will affect the accuracy of the SemCom model. 
% Intuitively, a higher compression rate will improve accuracy, but a lower compression rate will shorten the communication energy and latency.
Intuitively, a lower compression rate reduces the amount of transmitted data, lowering energy consumption and latency, but it also leads to information loss, which affects the accuracy of SemCom. In contrast, a higher compression rate preserves more semantic information but requires greater transmission power, increasing energy consumption and delay.
To effectively balance these trade-offs, resource allocation must jointly optimize key variables such as compression rate, transmission power, the binary ($0$-$1$) subcarrier allocation indicator, and computational frequency. Therefore, we propose a resource allocation problem that dynamically adjusts these parameters to achieve an optimal balance between model accuracy, energy efficiency, and latency.
Additionally, we have incorporated constraints to ensure that the optimization variables are in a feasible range.

Note that the scenario we discuss in this paper is mainly about mobile devices. Typically, mobile devices are not plentiful in computing and communication resources. Therefore, if the computation energy consumption of mobile devices needs to be minimized, when the joint optimization of communication energy or latency is not considered, the optimization problem is merely about how to allocate the CPU frequency, which must be the smaller, the better. However, for the operation of the system, this simple and rough allocation is unreasonable. Thus, we need to consider joint optimization to optimize latency, energy and accuracy together to ensure that the system's resource allocation is more holistic and reasonable.

\subsection{Contributions}
The contributions of this paper are as follows:\vspace{1pt} 
\begin{itemize}
    \item We are the first to propose an optimization problem for the FedSem system. We not only optimize the total energy and time consumption of the FL training process, but also optimize the transmission energy and the accuracy of SemCom.\vspace{1pt} 
    \item We devise a resource allocation algorithm that decomposes the original problem into two subproblems. For the \mbox{non-convex} problems, we transform them into convex forms by using convex techniques, ensuring tractability and solution effectiveness. Our approach includes a detailed analysis of time complexity and convergence, proving the practicality of the algorithm.\vspace{1pt} 
    \item Numerical experiments reveal superior performance of our algorithm in various conditions compared to baseline methods. The results validate the effectiveness of our solution in enhancing the operational efficiency of FedSem systems.
\end{itemize}

\subsection{Organization}
In Section \ref{sec:literature_review}, we discuss the literature review. In Section \ref{sec:sys_model}, we illustrate the system model and some related variables and equations. Section \ref{sec:joint_opt} introduces the joint optimization problem we formulate, and the corresponding solution is also provided. Time complexity and convergence are also discussed. Then, Section~\ref{sec:results} shows the numerical results. Finally, Section \ref{sec:conclusion} summarizes the whole paper.

\section{Literature Review} \label{sec:literature_review}
In this section, we discuss the related work in terms of two aspects: SemCom and FL. We also compare our work with other related work.

\subsection{Related Work on Semantic Communication}
% \textbf{2323}
Recently, SemCom has attracted more attention due to the quest for higher communication efficiency in the next generation of communication technologies \cite{wang2022transformer,pala2023spectral}. 
% \textbf{Applications on }
A number of works investigated the applications of SemCom on text \cite{xie2021deep,liu2022extended,peng2022robust}, images \cite{bourtsoulatze2019deep,jankowski2020wireless}, speech \cite{weng2021semantic} and videos \cite{jiang2022wireless}. 
% Furthermore, SemCom for some specific tasks, such as object recognition and scene classification tasks, semantic
% communications can significantly reduce the transmission overhead. 
Specifically, \cite{xie2021deep} presented DeepSC, a SemCom system using deep learning to optimize text transmission at the semantic level, focusing on sentence meaning rather than traditional error correction. It also utilized the Transformer model \cite{vaswani2017attention}, transfer learning \cite{weiss2016survey} for adaptability, and introduced a 'sentence similarity' metric for performance evaluation. \cite{liu2022extended} developed an Extended Context-based SemCom (ECSC) system that improves text transmission by utilizing context information within and between sentences for better accuracy, employing attention mechanisms and Transformer-XL at the encoder and decoder, respectively. R-DeepSC, proposed by \cite{peng2022robust}, introduced a robust deep learning-based system, which employed a calibrated self-attention mechanism and adversarial training to mitigate semantic noise, demonstrating superior performance over traditional models that only address physical noise.

SemCom can also extract information from multimedia data that is more complex than text. For image transmission, \cite{bourtsoulatze2019deep} introduced a deep learning-based joint source and channel coding (JSCC) technique that bypassed traditional compression and error correction codes, instead directly mapping pixels to channel symbols using two convolutional neural networks (CNNs) as encoder and decoder. \cite{jankowski2020wireless} used a deep neural network (DNN) for image compression and coding, while the analog scheme used joint source and channel coding (JSCC) for direct feature mapping to offer better accuracy and efficiency under variable channel conditions. \cite{weng2021semantic} presented DeepSC-S, a deep learning-enabled SemCom system for speech signals, utilizing a squeeze-and-excitation (SE) network and an attention mechanism. DeepSC-S was designed to handle varying channel conditions without retraining. \cite{jiang2022wireless} discussed a semantic video conferencing (SVC) system that sent key frames of videos, focusing on motions due to static backgrounds and infrequent speaker changes. It introduced an enhanced error management system, SVC-HARQ, with a new semantic error detector and an SVC-CSI for channel feedback, improving transmission efficiency and performance.

\textbf{Resource Allocation in SemCom}. There are also studies about resource allocation for SemCom \cite{yang2023semcomJSAC,li2023resource, liu2023semTCCN}. \cite{yang2023semcomJSAC} proposed an alternating method to calculate semantic information extraction ratio and computation frequency for minimizing transmission and computation energy. \cite{li2023resource} optimized the overall latency for a downlink SemCom system while guaranteeing the \mbox{physical-layer} security. \cite{liu2023semTCCN} introduced a resource allocation scheme to improve the task transmission probability while promising latency.

\subsection{Related Work on Federated Learning}
Due to limited communication and computation resources of local devices, FL usually takes longer to reach global convergence, affecting energy consumption and model performance.

\textbf{Resource Allocation in FL}. Related studies \cite{zhou2022joint,dinh2020federated,zhou2023resource,luo2020hfel,nguyen2020efficient,liu2022joint,chen2020joint,yang2020energy} have paid great attention to the resource allocation within the FL system. Specifically, \cite{zhou2022joint} proposed an alternating optimization-based resource allocation method to jointly optimize the energy and time consumption of an FL system, considering the transmission power, computational capability and bandwidth of local devices. A novel FL algorithm designed to manage heterogeneous data across diverse user equipment, FEDL, is formulated by \cite{dinh2020federated}. It optimizes the trade-off between data privacy and model performance. Furthermore, \cite{zhou2023resource} analyzed the relationship between object detection accuracy and image resolution, introducing an algorithm that optimizes energy and time without losing the performance of the global FL model.
Some papers also focused on the user (i.e., device/edge) association to reduce consumption in FL. \cite{luo2020hfel} introduced the HFEL framework, which combines edge association and resource scheduling to improve cost savings and training performance.
Similarly, a joint device scheduling and resource allocation method is presented by \cite{nguyen2020efficient}, running parallel stochastic gradient descent on a subset of the user equipment to optimize the model performance within a limited time budget. Besides, some works also considered the impact of user data distribution on resource allocation in FL, such as \cite{liu2022joint}. Moreover, \cite{chen2020joint} formulated a joint learning and communication framework, optimizing user selection and uplink resource allocation to improve FL accuracy under bandwidth limitations. \cite{yang2020energy} addressed energy-efficient FL by jointly optimizing computation and transmission resources, achieving significant energy savings. These studies highlight the growing focus on FL efficiency in real-world wireless environments.

\textbf{FL and SemCom}. Most studies \cite{tong2021federated, tongSemComJournal2021,loc2023federated,weiFedSem2023Letters,li2023CATFL,xie2023asynchronous,xing2023deepSemCom, xing2023fedDistillSC,liu2023SemCom2023VTC,song2023robustFedSem} related to the combination of FL and SemCom focused on devising new training strategies to improve the SemCom performance.
To further improve the accuracy of SemCom, \cite{tong2021federated} and \cite{tongSemComJournal2021} proposed an FL-based audio SemCom network, where a server and multiple devices collaboratively trained an autoencoder. Different from \cite{tong2021federated} and \cite{tongSemComJournal2021}, \cite{loc2023federated} proposed a new mechanism called FedLol, where the better-performed local models would be given more weights in aggregation rather than simply averaging the parameters from all participating models. 
% However, both \cite{tong2021federated} and \cite{loc2023federated} only proposed FL-based SemCom frameworks without considering how to optimize the considerable latency and energy consumption caused by the FL process.
One study \cite{weiFedSem2023Letters} investigated a new FedSem framework that trained semantic-channel encoders on local devices with the help of a semantic-channel decoder on the base station. 
Moreover, \cite{li2023CATFL} designed a trustworthy FL-based SemCom framework, which could authenticate client and server identities, through presenting a pseudonym generation strategy.
\cite{xie2023asynchronous} proposed an asynchronous FL framework to recognize multiple license plates through SemCom.
Besides, an FL-assisted multi-user deep SemCom system was proposed by \cite{xing2023deepSemCom} with dynamic model aggregation with the consideration of time-varying channel conditions. Apart from \cite{xing2023deepSemCom}, \cite{xing2023fedDistillSC} proposed a SemCom system assisted by federated distillation for image classification while also considering time-varying channel conditions.
\cite{liu2023SemCom2023VTC} devised an FL-assisted vehicular SemCom system and solved a semantic utility maximization problem by using deep reinforcement learning.
Additionally, \cite{song2023robustFedSem} designed a robust FL-assisted image SemCom framework that could defend against Byzantine attacks.

% of the above works do not consider optimizing the considerable delay and energy consumption caused by the FL process.

\subsection{Comparing Our Work with the Literature}
The previous related work that combined FL and SemCom \cite{tong2021federated, tongSemComJournal2021,loc2023federated,weiFedSem2023Letters,li2023CATFL,xie2023asynchronous,xing2023deepSemCom, xing2023fedDistillSC,song2023robustFedSem},\cite{10251890,10422907} mainly focused on proposing novel FedSem models, which is not the focus of this study.
These studies primarily explored the integration of FL with SemCom systems, focusing on various applications and optimization strategies. For instance, Tong et al. \cite{tong2021federated} proposed an FL-based audio semantic communication model to enhance audio transmission efficiency over wireless networks. Similarly, Nguyen et al. \cite{loc2023federated} introduced an efficient FL framework aimed at training SemCom systems, emphasizing the reduction of communication overhead during the training process. Wei et al. \cite{weiFedSem2023Letters} focused on federated semantic learning driven by information bottlenecks to optimize task-oriented communications. Additionally, Li et al. \cite{li2023CATFL} presented a certificateless authentication-based trustworthy FL approach designed for 6G SemCom systems.
Xu et al. \cite{10422907} focused on FL-powered SemCom for UAV swarm cooperation, highlighting the role of FL in enhancing collaborative capabilities among UAVs.
While these studies contribute significantly to the development of FedSem systems, they usually overlook the critical aspect of resource allocation within these systems. Our research addresses this gap by developing a novel resource allocation algorithm specifically designed for FL-enabled SemCom systems. This approach not only improves communication efficiency but also ensures optimal utilization of network resources, thereby improving overall system performance. By focusing on the intersection of resource allocation and the FedSem system, our work introduces a unique perspective that complements and extends the existing literature.

Only \cite{liu2023SemCom2023VTC} considered resource allocation in FL-empowered vehicular SemCom by focusing on semantic utility maximization. However, their approach did not account for essential factors such as energy consumption and model accuracy. In contrast, our research presents an optimization strategy for FL-assisted SemCom systems, aiming to concurrently minimize energy consumption and latency while improving model accuracy. This holistic approach ensures a more efficient and effective deployment of the FedSem system.
% considered resource allocation, but its problem was about semantic utility maximization without considering energy and model accuracy, while our paper considers optimizing energy, latency and accuracy of the FedSem system.
% Hence, we are the first to propose an optimization problem to minimize the energy, latency and maximize the accuracy of the FedSem system.

% \setlength{\abovedisplayskip}{3pt plus 2pt minus 0pt}
% \setlength{\belowdisplayskip}{3pt plus 2pt minus 0pt}
% \setlength\abovedisplayshortskip{3pt plus 2pt minus 0pt}
% \setlength\belowdisplayshortskip{2pt plus 2pt minus 0pt}
\section{System Model} \label{sec:sys_model}
\begin{figure*}
    \centering
    \includegraphics[width=0.8\linewidth]{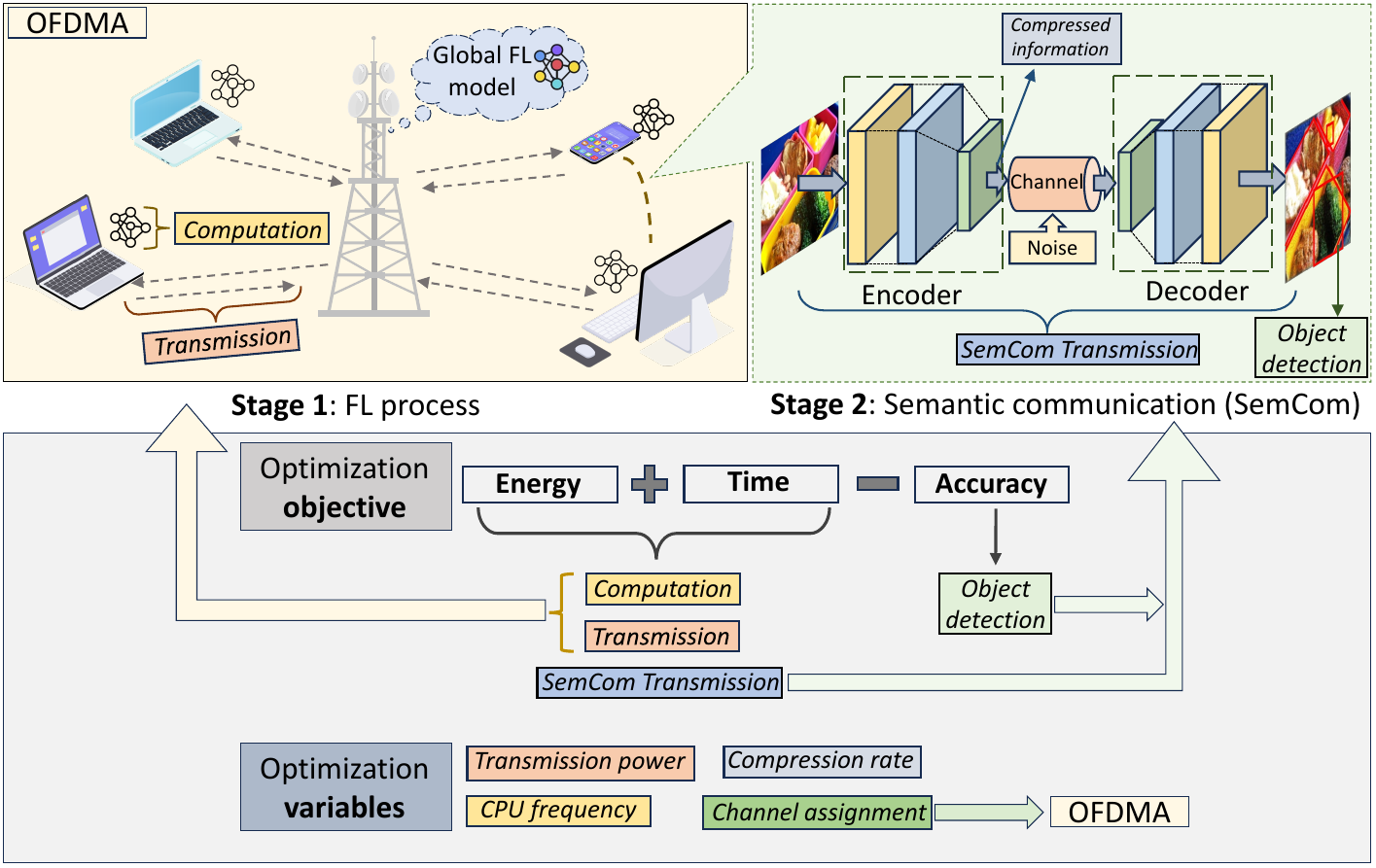}
    \caption{The overall system model and the optimization problem formulated in this paper.}
    \label{fig:system_model}
\end{figure*}

In this paper, we consider an FL-based SemCom system for image transmission. The channel access scheme is OFDMA. As shown in Fig. \ref{fig:system_model}, there are two stages: (1) \textbf{Stage 1}: The FL process. (2) \textbf{Stage 2}: The SemCom process. Next, we will introduce the two stages in detail.

\subsection{Stage 1: The Federated Learning Process}
In the FL process, multiple devices work collaboratively to train a global model, and the model adopts an autoencoder architecture (i.e., the encoder followed by the decoder). The encoder is parameterized by a convolutional neural network (CNN), and the same architecture is mirrored for the decoder. Between the encoder and the decoder, we add additive white Gaussian noise (AWGN) to make the model more robust to the practical scenario.
The loss function utilized on each local device (i.e., the metric used for assessing the autoencoder) is the mean squared error (MSE). The FL process is to minimize the MSE of the reconstructed images of the global model. 
Note that the performance of SemCom varies among different fading channels. 
However, we assume $t \in \{1,\dots,T\}$ represents the current timeslot. During the timeslot $t$, the channel is considered to exhibit block fading, meaning the channel state remains constant. The channel state varies across different timeslots. In this paper, we consider one timeslot $t$ instead of multiple timeslots, since we aim to establish a baseline performance to find out the relationship between the baseline SemCom performance and the compression rate $\rho$.
% The reason that we did not incorporate the specific fading is that we just aim to establish a baseline performance to find out the relationship between the baseline SemCom performance and the compression rate $\rho$. 
% \textcolor{red}{Note that the performance of SemCom varies across different fading channels due to channel-dependent distortions in semantic transmission. The reason we did not incorporate specific fading effects in our model is that our primary objective is to establish a baseline performance, allowing us to isolate and analyze the fundamental relationship between SemCom performance and the compression rate $\rho$. }
Focusing on this baseline scenario allows us to better understand how compression influences semantic accuracy and transmission efficiency before introducing additional complexities, such as dynamic fading effects, into the optimization framework.
Then, we can optimize the resource allocation of the system.
In our paper, we assume that the FL has entered a post-training stage. The trained model needs to be regularly updated for fine-tuning.

Assume there are $N$ mobile devices, one base station and $K$ subcarriers. We denote $\mathcal{N}:=\{1,\dots,N\}$ as the set of $N$ devices and $\mathcal{K}:=\{1,\dots,K\}$ as the set of $K$ subcarriers.
The total bandwidth is $B$. Besides, assume that one subcarrier is allocated to one user at most to prevent interference.
We define the transmission power of the \mbox{$n$-th} MAR device on the $k$-th subcarrier as $p_{n, k}$ and the channel gain as $g_{n,k}$.
Hence, the transmission rate of the \mbox{$n$-th} MAR device on the \mbox{$k$-th} subcarrier is
\begin{align}
    r_{n,k}(p_{n,k}) = \bar{B}\log_2(1+\frac{p_{n,k}g_{n,k}}{N_0\bar{B}}),
\end{align}
where $\bar{B} = \frac{B}{K}$, and $N_0$ is the noise spectral density. 
Moreover, the total transmission rate of the \mbox{$n$-th} device will be 
\begin{align}
    r_n = \sum_{k\in \mathcal{K}} x_{n,k}\cdot r_{n,k}(p_{n,k}),
\end{align}
where $\mathcal{K}=\{1,2,\cdots, K\}$, and $x_{n,k}\in \{0,1\}$ is the indicator that whether the \mbox{$k$-th} subcarrier is allocated to the \mbox{$n$-th} device. Also, the total transmission power of the \mbox{$n$-th} device is
\begin{align}
    p_n = \sum_{k\in \mathcal{K}}p_{n,k}.
\end{align}

% \textcolor{red}{No need to do the multiplication. As for transmission energy, $\frac{p_nD_n}{r_n}$, $p_n$ and $D_n$ are actually both fixed constants, what you are supposed to optimize is how to allocate $p_(n,k)$ to maximize $r_n$. According to (1) and (2), if $x_{n,k}$ is $0$, then corresponding $r_{n,k}(p_{n,k})$ does not contribute to $r_n$, so optimized $p_{n,k}$ would be $0$. The same is true for transmission delay.}

\textit{The Energy and Time Consumption: }
During this FL training process, the total energy consumption consists of the transmission energy and local computation energy.

For each device $n$, the transmission time (i.e., delay) is 
\begin{align}
   {\color{red}\tau}_n = \frac{D_n}{r_n},
\end{align}
where $D_n$ is the size of the model parameters uploaded by device $n$. Therefore, the transmission energy becomes
\begin{align}
    E_n^t = p_n\cdot {\color{red}\tau}_n.
\end{align}

The local computation time refers to the time used for training the encoder, and it is
\begin{align}
    t_n^c = \eta\frac{c_nd_n}{f_n},
\end{align}
where $\eta$ refers to the number of local iterations, $c_n$ means the CPU cycles per sample, $d_n$ is the number of samples on device $n$, and $f_n$ is the CPU frequency of device $n$. Then, the local computation energy is
\begin{align} \label{equa:FL-cmp-e}
    E_n^c = \xi \eta c_nd_nf_n^2,
\end{align}
where $\xi$ is the effective switched capacitance.

Note that we only consider the uplink process of FL in this paper. Therefore, the overall time consumption in FL becomes:
\begin{align} \label{def:T}
    T_{FL} = \max_{n\in \mathcal{N}} \{{\color{red}\tau}_n + t_n^c\},
\end{align}
where $\mathcal{N} = \{1,2,\cdots,N\}$. 

\subsection{Stage 2: The Semantic Communication Process}
After the FL process, the second stage is for the devices to utilize the trained encoders and decoders for SemCom.
% The scenario is that one device sends one image to another device, which needs to do object detection tasks.
Note that the utilized SemCom scheme follows the JSCC proposed by \cite{bourtsoulatze2019deep}. 
The target task of the SemCom stage in this paper is to transmit an image from a sender to a receiver through SemCom, and the receiver recovers the image the same as the original image as possible. Also, the receiver has an additional task---object detection. The reason that we adopt object detection is to further validate our FedSem system in a more practical way.
In the context of SemCom, the process of transmitting an image involves two fundamental stages: encoding and decoding. The semantic encoder first extracts the essential semantic features of the original image, and the channel encoder will compress the information. Subsequently, the compressed representation is sent through a communication channel to another device. Upon reception, the channel decoder first extracts the semantic features. The semantic decoder reconstructs the image from the semantic information and attempts to restore it to its original state.

However, there are several factors that can contribute to a reduction in image clarity and, consequently, a decline in object detection accuracy. They include the noise from the communication channel and inherent information loss associated with the compression step (i.e., the encoder compresses the original image). The noise could corrupt the transmitted data, and the compression step leads to a loss of some image details, further diminishing the clarity and fidelity of the reconstructed image. As a result, the final clarity of the reconstructed image might be lower than that of the original one, which influences the object detection accuracy. In this paper, we focus on the impact brought by the compression step on object detection accuracy. 

\subsubsection{The Accuracy Function}
We define the compression rate as $\rho$. The accuracy function is defined as $A_n(\cdot)$ with $\rho$ as the variable for each device. Note that we have assumed the FL has entered a post-training stage.
The term $A_n(\rho)$ reflects the empirically observed performance of the well-trained model under varying compression rates $\rho$. 
Then, the trained SemCom models achieve a stable performance that can be approximated as a function of the compression rate $\rho$.
In this paper, we utilize object detection accuracy as an auxiliary metric to assess the performance of SemCom. 
To find the relationship between the model accuracy and $\rho$, we run the YOLOv3~\cite{redmon2018yolov3} algorithm by using the COCO dataset~\cite{coco_dataset} under different $\rho$.
Note that for the sake of clarity and reader comprehension, the model `accuracy' we refer to is mAP (mean average precision), which is a common metric used in the computer vision research community for assessing the performance of object detection models.

\begin{figure}
    \centering
    \includegraphics[width=0.3\textwidth]{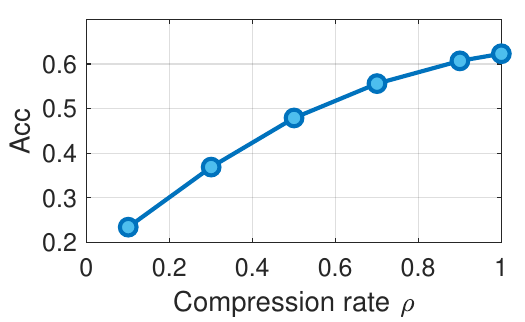}
    \caption{The accuracy versus $\rho$.\vspace{-10pt}}
    \label{fig:acc_snr}
\end{figure}

As shown in Fig. \ref{fig:acc_snr}, the curve is increasing and concave. 
Hence, we suppose that the accuracy function is non-decreasing and concave. Also, we provide this assumption:
\begin{assumption}{1}{}
   The accuracy function $A_n(\rho)$ is \mbox{non-linear}, increasing and concave, $\forall~n\in\mathcal{N}$, $\rho \in [0,1]$.
\end{assumption}

\subsubsection{The Energy and Time Consumption}
For Stage 2, we exclusively take into account the transmission energy and latency.
As previously mentioned, during Stage 2, we only assess the impact of compression rate, $\rho$, whereas changes in $\rho$ have a relatively minor effect on computation.
Hence, we solely consider the impact of $\rho$ on the communication overhead for SemCom, as our primary focus is to optimize the effect that training a SemCom model via FL will have on subsequent semantic communications.
 The primary influence of $\rho$ pertains more to the size of the transmitted compressed information. Thus, our analysis is primarily centered on its effect on transmission energy and time consumption.

We assume that there are $L$ rounds of semantic communications for each device $n$. Hence, at the $l$-th round, the SemCom transmission time of the device $n$ is given as
\begin{align}
    T_{n,l}^{sc} = \frac{\rho\mathcal{C}_{n,l}}{r_n},~\forall l \in \{1,\dots,L\},
\end{align}
where $\mathcal{C}_{n,l}$ is the size of semantic compressed information of each device $n$ at the $l$-th round.
Then, the total SemCom transmission time for $L$ communication rounds will be
\begin{align}
    T_{n}^{sc} = \sum_{l=1}^L T_{n,l}^{sc} =  \frac{\rho \mathcal{C}_n}{r_n},
\end{align}
where $\mathcal{C}_n = \sum_{l=1}^L \mathcal{C}_{n,l}$.

Besides, the SemCom transmission energy of the device $n$ at the $l$-th round is
\begin{align}
    E_{n,l}^{sc} = p_n\cdot\frac{\rho\mathcal{C}_{n,l}}{r_n},~\forall l \in \{1,\dots,L\}.
\end{align}
The total SemCom transmission energy becomes
\begin{align}
    E_{n}^{sc} = \sum_{l=1}^L E_{n,l}^{sc} = p_n\cdot\frac{\rho \mathcal{C}_n}{r_n}.
\end{align}

As for the size (i.e., $\mathcal{C}_{n,l}$) of semantic compressed information of each device at the $l$-th round, we provide an assumption:
\begin{assumption}{2}{}
     The size, $\mathcal{C}_{n,l}$, of semantic compressed information is known in advance or the estimation of $\mathcal{C}_{n,l}$ can be useful for the system design.
\end{assumption}

The reason that we give this assumption is that $\mathcal{C}_{n,l}$ can be calculated based on the SemCom model (e.g., the autoencoder), which could be utilized in our optimization problem.
Therefore, we assume that the size of semantic compressed information will be known in advance. 
 Or, our estimation of $\mathcal{C}_{n,l}$ can be useful for the system design.

\section{Joint optimization of energy, time and accuracy} \label{sec:joint_opt}
In this section, we formulate an optimization problem to optimize the weighted sum of FL energy (i.e., $\sum_{n\in\mathcal{N}}E_n^t+E_n^c$), SemCom transmission energy (i.e., $\sum_{n\in\mathcal{N}}E_n^{sc}$), and FL time consumption (i.e, $T_{FL}$) as well as the model accuracy (i.e., $\sum_{n\in\mathcal{N}}A_n(\rho)$).

We present the problem as follows.

\noindent
\textbf{Problem}~$\mathbb{P}_1$:
\begin{subequations} \label{problem:original}
\begin{align}
    \min_{\boldsymbol{f},\boldsymbol{P}, \boldsymbol{X}, \rho} & \kappa_1\sum_{n\in\mathcal{N}} E_n \!+\! \kappa_2 T_{FL} \!-\! \kappa_3 \sum_{n\in\mathcal{N}} A_n(\rho), \tag{\ref{problem:original}}\\
    \text{Subject to,} &  ~p_{n,k} \leq x_{n,k}P_{n}^{max},~\forall n \in \mathcal{N},~\forall k \in \mathcal{K}, \label{constra:pmax} \\
    & p_n \leq P_{n}^{max},~\forall n \in \mathcal{N}\label{constra:sum_pnk_max}\\
    & f_n \leq f_n^{max},~\forall n \in \mathcal{N}, \label{constra:fmax}\\
    % & r_n \geq r^{min}, \label{constra:rmin} \\
    & \sum_{n=1}^N x_{n,k} \leq 1,~\forall k \in \mathcal{K}, \label{constra:sum_xnk} \\
    & x_{n,k} \in \{0, 1\},~\forall n \in \mathcal{N}, \forall k \in \mathcal{K}, \label{constra:0_1_xnk}\\
    & T_{n}^{sc} \leq T_{n,max}^{sc},~\forall~n\in\mathcal{N}, \label{constra:T_secom_limit} \\
    & \rho \leq 1, \label{constra:rho_range}
\end{align}
\end{subequations}
where $\boldsymbol{f}$, $\boldsymbol{P}$, $\boldsymbol{X}$ and $\rho$ are optimization variables, and $E_n=E_n^t + E_n^c + E_n^{sc},\forall n\in\mathcal{N}$. $\boldsymbol{f}:=[f_1,\dots, f_N]$ is a vector containing the computational frequency of each device $n$. $\boldsymbol{P}:=[p_{n,k}]_{N\times K}$ ($n\in\mathcal{N},k\in\mathcal{K}$) is a matrix where each entry is the transmission power of the $n$-th device on the $k$-th subcarrier, and $\boldsymbol{X}:=[x_{n,k}]_{N\times K}$ ($n\in\mathcal{N},k\in\mathcal{K}$) is the indicator matrix representing whether the $k$-th subcarrier is allocated to the $n$-th device. $\rho$ is the compression rate of the SemCom model.
Besides, $\kappa_1$, $\kappa_2$, and $\kappa_3$ are \mbox{non-negative} weight parameters, corresponding to energy, time, and accuracy, respectively. 
The unit of the weight parameter $\kappa_1$ for the energy consumption is the inverse of Joule (i.e., J$^{-1}$), and the unit of the weight parameter $\kappa_2$ for the delay is the inverse of second (i.e., s$^{-1}$). Also, $\kappa_3$ of accuracy is just a unitless value since accuracy is a proportion or percentage.
We can adjust each weight parameter to tune the focus of the optimization problem. For instance, we could increase $\kappa_1$ to make the optimization problem more focused on optimizing energy.
Constraints (\ref{constra:pmax}) and (\ref{constra:sum_pnk_max}) set the range of the transmission power $p_{n,k}$ and $p_n$. Constraint (\ref{constra:fmax}) limits the range of computational frequency $f_n$. Additionally, constraints (\ref{constra:sum_xnk}) and (\ref{constra:0_1_xnk}) ensure that at most one subcarrier is allocated to one user and $x_{n,k}$ is either $0$ or $1$. Constraint (\ref{constra:T_secom_limit}) limits the maximum transmission time of SemCom for each device. In addition, constraint (\ref{constra:rho_range}) sets the range of compression rate in $[0,1]$.

% \textcolor{red}{If you only consider the uplink progress, the constraint (\ref{constra:pmax}) may should be written as $\sum_{k=1}^K p_{n,k} \leq P_n^{max}$ (i.e., the sum of power allocated to each subcarrier by user $n$ cannot exceed the upper limit).}

% The complete form of (\ref{problem:original}) is as follows:
% \begin{align}
%     &\min_{p_{n,k}, f_{n}, x_{n,k}, \rho}\notag\\
%     & \kappa_1 \sum_{n\in \mathcal{N}} \Big( \frac{(\sum_{k \in \mathcal{K}}p_{n,k}) D_n}{\sum_{k \in \mathcal{K}}x_{n,k}\bar{B}\log_2(1+\frac{p_{n,k}g_{n,k}}{N_0 \bar{B}})} \notag + \xi \eta c_nd_nf_n^2 \notag \\
%     &+ \frac{(\sum_{k \in \mathcal{K}}p_{n,k}) \mathcal{C}}{\sum_{k \in \mathcal{K}}x_{n,k}\bar{B}\log_2(1+\frac{p_{n,k}g_{n,k}}{N_0 \bar{B}})} \Big)\notag\\
%     & + \kappa_2 \max_{n\in \mathcal{N}}\{\frac{D_n}{\sum_{k \in \mathcal{K}}x_{n,k}\bar{B}\log_2(1+\frac{p_{n,k}g_{n,k}}{N_0 \bar{B}})} + \eta \frac{c_n d_n}{f_n}\} \notag\\
%     &-\kappa_3 \sum_{n\in\mathcal{N}} A_n(\rho).
% \end{align}

Since $x_{n,k}$ is a binary value, the optimization problem (\ref{problem:original}) is obviously non-convex. 
Problem $\mathbb{P}_1$ contains a term $\sum_{n\in\mathcal{N}} E_n^t$, and the complete form is
$$ \sum_{n\in\mathcal{N}} \frac{(\sum_{k\in\mathcal{K}}p_{n,k})D_n}{\bar{B}\log_2(1+\frac{p_{n,k}g_{n,k}}{N_0\bar{B}})},$$
where $p_{n,k}$ is the optimization variable. 
Considering the numerator is non-negative and convex, and the denominator is positive and concave, $E_n^t$ is pseudoconvex (referring to Theorem 3.2.10 of \cite{cambinigeneralized}). However, $\sum_{n\in\mathcal{N}}E_n^t$ is a sum-of-ratios form, which is NP-complete \cite{jong2012efficient}, complicating the \mbox{non-convex} problem. Ideally, we want to find an approach to find a global optimum for the objective function of $\mathbb{P}_1$. However, since $\mathbb{P}_1$ has four optimization variables, and three of them are multi-dimensional, it is quite difficult to directly obtain the global optimal solution in its current form. 
Hence, we need to transform this problem into a solvable and simpler form.

% {\color{red}First of all, we relax $x_{n,k}$ into a continuous variable by adding a constraint $\sum_{n\in \mathcal{N}}\sum_{k\in \mathcal{K}} x_{n, k}(1-x_{n,k}) \leq 0$. Besides, in order to make constraint (\ref{constra:pmax}) tighter, we adopt the method from \cite{nam2019ofdma} to replace (\ref{constra:pmax}) by
% $$
%     p_{n,k} \leq (x_{n,k})^q P_{n}^{max},~\forall n \in \mathcal{N},~\forall k \in \mathcal{K},
% $$
% where $q\geq 1$ and $q$ is an integer.
% Note that for $x_{n,k}\in\{0,1\}$, $x_{n,k}=(x_{n,k})^q$. Since we relax $x_{n,k}$ into a continuous variable in the interval $[0,1]$, $x_{n,k}\geq(x_{n,k})^q$. Thus, constraint (\ref{constra:pmax}) becomes tighter if it is replaced by the above constraint.}

First, we add one auxiliary variable $\mathcal{T}$ to transform the max function $T_{FL}$ (\ref{def:T}) with the new constraint $\tau_n + t_n^c \leq \mathcal{T}$, and transform the original optimization problem into an epigraph form. 
The epigraph form helps to transform the max function into a linear objective $\mathcal{T}$, which allows us to convert the max function into a more tractable form.
Then, the problem becomes

\setlength{\abovedisplayskip}{4pt plus 1pt minus 0pt}
\setlength{\belowdisplayskip}{4pt plus 1pt minus 0pt}
\setlength\abovedisplayshortskip{2pt plus 1pt minus 0pt}
\setlength\belowdisplayshortskip{2pt plus 1pt minus 0pt}

\noindent
\textbf{Problem}~$\mathbb{P}_2$:
\begin{subequations} \label{problem:relax_version}
  \begin{align} 
~     \min_{\boldsymbol{f}, \boldsymbol{P}, \boldsymbol{X}, \rho, \mathcal{T}} & \kappa_1\sum_{n\in\mathcal{N}} E_n \!+\! \kappa_2 \mathcal{T} \!-\! \kappa_3 \sum_{n\in\mathcal{N}} A_n(\rho), \tag{\ref{problem:relax_version}} \\
    \text{Subject to}, &~(\ref{constra:pmax})\text{--}(\ref{constra:rho_range}),\notag\\
    & \tau_n + t_n^c \leq \mathcal{T},~\forall n \in \mathcal{N}. \label{constra:max_T}
\end{align}  
\end{subequations}

% {\color{red}
% \begin{subequations} \label{problem:relax_version}
%   \begin{align} 
%     \min_{p_{n,k}, f_n, x_{n,k}, \rho} & \kappa_1\sum_{n\in\mathcal{N}} (E_n^t \!+\! E_n^c \!+\! E_n^{sc}) \!+\! \kappa_2 \mathcal{T} \!-\! \kappa_3 \sum_{n\in\mathcal{N}} A_n(\rho), \tag{\ref{problem:relax_version}} \\
%     \text{Subject to}, &~(\ref{constra:fmax}), (\ref{constra:sum_xnk}), (\ref{constra:T_secom_limit}), \notag
% \end{align}  
% \end{subequations}

% }

% However, the above problem is still intractable due to its complex form and non-convexity. Hence, we continue transforming the current problem.

% Therefore, Problem (\ref{problem:relax_version}) is divided into two subproblems based on the degree of association among the variables, as shown below: 

However, the above problem $\mathbb{P}_2$ is still intractable due to its complex form and non-convexity. 
Hence, we devise an iterative optimization scheme to solve $\mathbb{P}_2$ by alternately optimizing two different sets of variables. Each iteration comprises two optimization stages below:
\begin{itemize}
    \item Step 1: Given fixed $(\boldsymbol{P},\boldsymbol{X})$, optimize $(\boldsymbol{f},\rho,\mathcal{T})$.
    \item Step 2: Given obtained  $(\boldsymbol{f},\boldsymbol{g},\boldsymbol{\lambda})$, optimize $(\boldsymbol{p},\boldsymbol{B})$.
\end{itemize}

According to the devised solution above, we are able to further decompose problem $\mathbb{P}_2$ in each step based on the degree of association among the variables. Problem decomposition is necessary when dealing with complex optimization problems that have multiple variables or constraints. Decomposing the problem into smaller, more manageable sub-problems also helps us apply specialized solutions to different parts of the problem. 
Given fixed $(\boldsymbol{P},\boldsymbol{X})$, $\mathbb{P}_2$ could be reformulated into $\mathbb{P}_3(\boldsymbol{f}, \rho, \mathcal{T})$, expressed as follows:

\textbf{Problem}~\textbf{$\mathbb{P}_3(\boldsymbol{f}, \rho, \mathcal{T})$}:
    \begin{subequations} \label{problem:sp1}
        \begin{align}
             \min_{\boldsymbol{f}, \rho, \mathcal{T}} & ~\kappa_1\sum_{n=1}^N (E_n^c + E_n^{sc})+ \kappa_2 \mathcal{T} - \kappa_3\sum_{n\in\mathcal{N}}A_n(\rho), \tag{\ref{problem:sp1}} \\
             \text{Subject to}, &~ (\ref{constra:fmax}), (\ref{constra:T_secom_limit}), (\ref{constra:rho_range}), (\ref{constra:max_T}), \notag
        \end{align}
    \end{subequations}
    
Similarly, given obtained $(\boldsymbol{f}, \rho, \mathcal{T})$, $\mathbb{P}_2$ could be simplified into the following problem $\mathbb{P}_4(\boldsymbol{P}, \boldsymbol{X})$:

\textbf{Problem $\mathbb{P}_4(\boldsymbol{P}, \boldsymbol{X})$}:
    \begin{subequations} \label{problem:sp2}
        \begin{align}
            \min_{\boldsymbol{P}, \boldsymbol{X}} &~\kappa_1\sum_{n=1}^N E_n^t+E_n^{sc}, \tag{\ref{problem:sp2}} \\
             \text{Subject to}, &~(\ref{constra:pmax}), (\ref{constra:sum_pnk_max}), (\ref{constra:sum_xnk})\text{--}(\ref{constra:T_secom_limit}), (\ref{constra:max_T}).\notag
        \end{align}
    \end{subequations}

Note that $\mathbb{P}_3(\boldsymbol{f}, \rho, \mathcal{T})$ is easy to solve due to its convexity, whereas $\mathbb{P}_4(\boldsymbol{P}, \boldsymbol{X})$ is not. Therefore, we will focus on elaborating how to solve $\mathbb{P}_4(\boldsymbol{P}, \boldsymbol{X})$. However, before addressing $\mathbb{P}_4(\boldsymbol{P}, \boldsymbol{X})$, we will first introduce how to solve $\mathbb{P}_3(\boldsymbol{f}, \rho, \mathcal{T})$.

\subsection{The solution to $\mathbb{P}_3(\boldsymbol{f}, \rho, \mathcal{T})$}

For $\mathbb{P}_3$, through observation, we could combine constraints (\ref{constra:T_secom_limit}) and (\ref{constra:rho_range}) as one: $$\rho \leq \rho^{max},$$
where $\rho^{max} = \min \big(1, \frac{T_{n,max}^{sc}r_n}{{\mathcal{C}_n}}|_{n\in\mathcal{N}}\big)$.

The subproblem $\mathbb{P}_3(\boldsymbol{f}, \rho, \mathcal{T})$ is convex, and thus it is easy to solve. We can apply Karush-Kuhn-Tucker (KKT) conditions to it. First, we write the Lagrange function: 
\begin{align}
    &L_1(\boldsymbol{f}, \mathcal{T}, \rho, \boldsymbol{\mu}, \boldsymbol{\upsilon}) = \kappa_1 \sum_{n=1}^N (\xi\eta c_nd_n f_n^2+p_n\frac{\rho \mathcal{C}_n}{r_n}) \notag \\
    & + \kappa_2 \mathcal{T}-\kappa_3\sum_{n=1}^NA_n(\rho) + \sum_{n=1}^N \mu_n\cdot(\tau_n+\eta \frac{c_nd_n}{f_n}-\mathcal{T}) \notag\\
    &+ \sum_{n=1}^N \delta_n\cdot(f_n-f_n^{max}),
\end{align}
where $\boldsymbol{\mu}:=[\mu_1,\dots,\mu_N]$ and $\boldsymbol{\delta}:=[\delta_1,\dots,\delta_N]$ are \mbox{non-negative} Lagrange multipliers.
After applying KKT conditions, we get

\textbf{Stationarity:}
\begin{align}
   & \frac{\partial L_1}{\partial f_n} = 2\kappa_1\xi\eta c_nd_nf_n-\frac{\mu_n\eta c_nd_n}{f_n^2} + \delta_n=0, \label{kkt:partial_fn}\\
  &  \frac{\partial L_1}{\partial \mathcal{T}} = \kappa_2-\sum_{n=1}^N\mu_n=0,\label{kkt:partial_T}\\
   & \frac{\partial L_1}{\partial \rho} = \sum_{n=1}^N (\frac{\kappa_1 p_n\mathcal{C}_n}{r_n}-\kappa_3\frac{\partial A_n(\rho)}{\partial \rho})=0, 
    \label{kkt:partial_rho}
    \end{align}

\textbf{Complementary slackness:}
    \begin{align}
    &\mu_n\cdot (\tau_n + \eta \frac{c_nd_n}{f_n}-\mathcal{T})=0, \label{kkt:mu_complementary_slackness}\\
    &\delta_n \cdot (f_n - f_n^{max}) = 0, \label{kkt:fn_complementary_slackness}
    \end{align}
    
\textbf{Primal feasibility:}
    \begin{align}
    &\text{(\ref{constra:fmax}), (\ref{constra:T_secom_limit}), (\ref{constra:rho_range}), (\ref{constra:max_T})},\notag
    \end{align}

\textbf{Dual feasibility:}
    \begin{align}
    &\mu_n, \delta_n \geq 0, \forall~n\in\mathcal{N}.
\end{align}

% Additionally, from conditions (\ref{kkt:partial_rho}) and (\ref{kkt:rho_complementary_slackness}), we could derive how to calculate $\rho$. If $\upsilon \neq 0$, $\rho = \frac{\textcolor{blue}{T_{n,max}^{sc}}r_n}{\mathcal{C}}$. However, since $\rho$ is univariate, it cannot satisfy all $r_n|_{n=1,\dots,N}$. 
% Thus, when $n=1\dots N$, only a unique $n$ can satisfy $\upsilon \neq 0$ and $\rho=\frac{\textcolor{blue}{T_{n,max}^{sc}}r_n}{\mathcal{C}}$.
% While for other values of $n$, it must be ensured that $\upsilon=0$. Given the calculated $\rho$, we can derive the value of $\upsilon$ using the bisection method based on (\ref{kkt:partial_rho}). 
% Once $\upsilon>0$, the calculated $\rho$ is valid. Otherwise, try another value of $n$, and repeat the same steps for computation. 
% Consequently, 
% \begin{equation} \label{rho_representation}
% \rho \!=\! \left\{
%     \begin{aligned}
%      &\min_n(\frac{\textcolor{blue}{T_{n,max}^{sc}}r_n}{\mathcal{C}}),~\text{if}~ \upsilon \!>\! 0~\text{and}~\upsilon_{j\neq n}=0, \forall~n,j \!\in\!\mathcal{N}, \\&\text{when using the}~\text{bisection method based on (\ref{kkt:partial_rho})}
% \end{aligned}
% \right.
% \end{equation}

Additionally, from (\ref{kkt:partial_rho}), we could use the bisection method to find the optimal $\rho$ given $\frac{\partial A_n(\rho)}{\partial \rho}$ is an increasing function. Based on (\ref{kkt:partial_rho}), we define a new function $\Delta(\rho)=\sum_{n=1}^N (\frac{\kappa_1 p_n\mathcal{C}_n}{r_n}-\kappa_3\frac{\partial A_n(\rho)}{\partial \rho})$.
Then, the optimal 
\begin{equation}
    \rho^* = \min\big(\rho^\#, \rho^{max} \big), \label{optimal_rho}
\end{equation}
where $\rho^\#$ satisfies $\Delta(\rho^\#)=0$.
% Let $\rho^*$ denote the calculated $\rho$.
Until now, we have derived the optimal $\rho^*$. 

% With the conditions (\ref{kkt:partial_fn}) and (\ref{kkt:partial_T}), it could be derived that
% \begin{align}
%     f_n=\sqrt[3]{\frac{\mu_n}{2\kappa_1\xi}},~ \sum_{n=1}^N \mu_n = \kappa_2. \label{kkt:fn_mun}
% \end{align}
% From (\ref{kkt:fn_mun}), we find the relationship between the computation frequency $f_n$ and the Lagrange multiplier $\mu_n$. Since $f_n$ cannot be $0$, we could also conclude $\mu_n>0$. Thus, it could be derived from (\ref{kkt:mu_complementary_slackness}) that
% \begin{align}
%     f_n = \frac{\eta c_nd_n}{\mathcal{T}-{\color{red}\tau}_n}. \label{fn_T}
% \end{align}
% Moreover, given $f_n=\sqrt[3]{\frac{\mu_n}{2\kappa_1\xi}}$ and $\sum_{n=1}^N \mu_n = \kappa_2$ in (\ref{kkt:fn_mun}), we could use the bisection method to calculate the value of $\mathcal{T}$ by
% \begin{align}
%     \sum_{n=1}^N \bigg[2\kappa_1 \xi \bigg(\min\Big(\frac{\eta c_nd_n}{\mathcal{T}-{\color{red}\tau}_n}, f_n^{max}\Big)\bigg)^3\bigg] =\kappa_2. \label{T_k2}
% \end{align}

From (\ref{kkt:partial_fn}), we could conclude that $\mu_n>0$ must hold; otherwise, (\ref{kkt:partial_fn}) is not valid.

Therefore, there will be two cases for Lagrange multipliers $\mu_n$ and $\delta_n$: 1) $\mu_n>0, \delta_n =0$; 2) $\mu_n>0, \delta_n >0$.

\textbf{Case 1} [$\mu_n>0, \delta_n =0$]:
With the conditions (\ref{kkt:partial_fn}) and (\ref{kkt:partial_T}), it could be derived that
\begin{align}
    f_n=\sqrt[3]{\frac{\mu_n}{2\kappa_1\xi}},~ \sum_{n=1}^N \mu_n = \kappa_2. \label{kkt:fn_mun}
\end{align}
From (\ref{kkt:fn_mun}), we find the relationship between the computation frequency $f_n$ and the Lagrange multiplier $\mu_n$. Thus, it could be derived from (\ref{kkt:mu_complementary_slackness}) that
\begin{align}
    f_n = \frac{\eta c_nd_n}{\mathcal{T}-\tau_n}. \label{fn_T}
\end{align}

We need to make sure that this result satisfies the primary feasibility. Hence, if (\ref{constra:fmax}) holds, this case is valid. Otherwise, $\delta_n =0$ is not valid, and $\delta_n >0$. Then, we could jump to Case 2.

\textbf{Case 2} [$\mu_n>0, \delta_n >0$]: Since $\delta_n>0$, it is derived from (\ref{kkt:fn_complementary_slackness}) that $f_n=f_n^{max}$.

From the above two cases, we could summarize that the optimal $\boldsymbol{f}^*$ is
\begin{align}
    f_n^* = \min\Big(\frac{\eta c_nd_n}{\mathcal{T}-\tau_n}, f_n^{max}\Big),~n\in\mathcal{N}.
\end{align}

Up to this step, we still have $\mathcal{T}$ left to derive.
Given $f_n=\sqrt[3]{\frac{\mu_n}{2\kappa_1\xi}}$ and $\sum_{n=1}^N \mu_n = \kappa_2$ in (\ref{kkt:fn_mun}), we could use the bisection method to calculate the value of $\mathcal{T}$ by
\begin{align}
    \sum_{n=1}^N \bigg[2\kappa_1 \xi \bigg(\min\Big(\frac{\eta c_nd_n}{\mathcal{T}-\tau}_n, f_n^{max}\Big)\bigg)^3\bigg] =\kappa_2. \label{T_k2}
\end{align}

To compute $\mathcal{T}$ from (\ref{T_k2}) by using the bisection method,
we first define a function $F(\mathcal{T})$ by moving everything to one side:
\[
F(\mathcal{T}) = \sum_{n=1}^N \Big[2\,\kappa_1\,\xi \,\big(\min\big(\frac{\eta\,c_n\,d_n}{\mathcal{T}-\tau_n},\,f_n^{max}\big)\big)^3\Big] 
\;-\;\kappa_2.
\]
We then find $\mathcal{T}$ such that $F(\mathcal{T}) = 0$. Because each term is continuous for $\mathcal{T} > \max(\tau_n)$ (the only complication is the piecewise nature of the $\min$ function, but it does not introduce jumps), we can apply the bisection method. Specifically, we choose two points $a$ and $b$ such that $F(a)$ and $F(b)$ have opposite signs; this ensures that a root lies between them. We repeatedly halve the interval $[a, b]$ by checking the sign of $F$ at the midpoint and then narrowing down the sub-interval that still contains the root. This process converges once the interval is sufficiently small, and the midpoint of the final interval is taken as the solution for $\mathcal{T}$.

Therefore, we provide the following theorem to solve $\mathbb{P}_3$.
\begin{theorem} \label{theorem:cal-f-rho}
    The optimal solution ($\boldsymbol{f}^*$, $\rho^*$ and $\mathcal{T}^*$) of $\mathbb{P}_3$ could be derived from
    \begin{align}
        f_n^* &= \min\Big(\frac{\eta c_nd_n}{\mathcal{T^\#}-\tau_n}, f_n^{max}\Big),~n\in\mathcal{N}, \\
        \rho^* &= \text{(\ref{optimal_rho})},\nonumber\\
        \mathcal{T}^* &= \max\Big(\tau_n + \frac{\eta c_nd_n}{f_n^*}|_{n\in\mathcal{N}} \Big),
    \end{align}
    where $\mathcal{T}^\#$ satisfies (\ref{T_k2}).
\end{theorem}

\subsection{The solution to $\mathbb{P}_4(\boldsymbol{P}, \boldsymbol{X})$}
% The difficulty of solving SP2 is the existence of $x_{n,k}p_{n,k}$.
% It is obvious that $x_{n,k}p_{n,k}$ is not convex nor concave. Motivated by \cite{zhao2023human}, $x_{n,k}p_{n,k}$ satisfies the following inequality 
% $$x_{n,k}p_{n,k} \leq x_{n,k}^2c_{n,k}+\frac{p_{n,k}^2}{4c_{n,k}},$$
% and the equality is obtained when $c_{n,k}=\frac{p_{n,k}}{2x_{n,k}}$. Therefore, we could replace $x_{n,k}p_{n,k}$ in problem (\ref{problem:sp2}) by $x_{n,k}^2c_{n,k}+\frac{p_{n,k}^2}{4c_{n,k}}$, and let $c_{n,k}=\frac{p_{n,k}}{2x_{n,k}}$.
% For ease of representation, we use 
% \begin{align}
%     G_n(x_{n,k},p_{n,k})= (\sum_{k\in\mathcal{K}}p_{n,k})D_n
% \end{align}
% in the rest of this paper.

% Nevertheless, due to $s_n=\sum_{n=1}^N \frac{\sum_{k=1}^K x_{n,k}\frac{p_{n,k}g_{n,k}}{N_0\bar{B}}}{N}$, $A_n(s_n)$ is not convex. However, we cannot transform the term $x_{n,k}p_{n,k}$ into $x_{n,k}^2c_{n,k}+\frac{p_{n,k}^2}{4c_{n,k}}$, which is convex, because the nature of the transformation cannot make the problem easier. According to the composition rule (Eq. (3.11) in \cite{boyd2004convex}), if $A_n(\cdot)$ is concave and non-decreasing and $s_n$ is concave, $A_n(s_n)$ is concave. Hence, we take the method from \cite{shen2018fractional} to transform $x_{n,k}p_{n,k}$ into $2z_{n,k}\sqrt{x_{n,k}}-z_{n,k}^2\frac{1}{p_{n,k}}$, which is concave, by introducing a new auxiliary variable $z_{n,k}=\sqrt{x_{n,k}}p_{n,k}$.

Now, we have yet to resolve problem $\mathbb{P}_4(\boldsymbol{P}, \boldsymbol{X})$, which is described as follows:
\begin{subequations} \label{problem:SP2_transformation1}
    \begin{align}
        \min_{\boldsymbol{P}, \boldsymbol{X}} &~\kappa_1\sum_{n\in\mathcal{N}} \frac{(\sum_{k\in\mathcal{K}}p_{n,k})\cdot(D_n+\rho \mathcal{C}_n)}{r_n}, \tag{\ref{problem:SP2_transformation1}}\\
        \text{Subject to,} &~ (\ref{constra:pmax}), (\ref{constra:sum_pnk_max}), (\ref{constra:sum_xnk})\text{--}(\ref{constra:T_secom_limit}), (\ref{constra:max_T}). \notag
    \end{align}
\end{subequations}
Since $x_{n,k}$ is a binary variable, $\mathbb{P}_4(\boldsymbol{P}, \boldsymbol{X})$ is complex and difficult to solve. 
Intuitively, we relax $x_{n,k}$ into a continuous variable in the $[0, 1]$ interval. 
This relaxation converts the problem from a \mbox{mixed-integer} programming problem to a continuous optimization problem, which is generally more tractable.
However, relaxing $x_{n,k}$ to the continuous domain might weaken the original constraint (\ref{constra:pmax}) (i.e., $p_{n,k} \leq x_{n,k}P_{n}^{max}$), making the solution less precise or meaningful.
To address this, a transformation is introduced to tighten the constraint and ensure it better approximates the original binary formulation.
We then adopt the method from \cite{nam2019ofdma} to replace (\ref{constra:pmax}) by
$$
    p_{n,k} \leq (x_{n,k})^q P_{n}^{max},~\forall n \in \mathcal{N},~\forall k \in \mathcal{K},
$$
where $q\geq 1$ and $q\in\mathbb{N}^+$.
Note that if $x_{n,k}\in\{0,1\}$, $x_{n,k}=(x_{n,k})^q$; if $x_{n,k}$ is a continuous variable in the interval $[0,1]$, $x_{n,k}\geq(x_{n,k})^q$. Thus, constraint (\ref{constra:pmax}) becomes tighter if it is replaced by the above constraint.

Additionally, to ensure that $x_{n,k}$ ultimately takes on a value of either $0$ or $1$, a constraint $\sum_{n\in \mathcal{N}}\sum_{k\in \mathcal{K}} x_{n, k}(1-x_{n,k}) \leq 0$ is introduced to replace the constraint (\ref{constra:0_1_xnk}) (i.e., $x_{n,k}\in\{0,1\}$). In light of $0\leq x_{n,k}\leq 1$, this constraint ensures that $x_{n,k}=0$ or $1$. 

In addition, the constraint (\ref{constra:T_secom_limit}) can be represented as $r_n\geq \frac{\rho\mathcal{C}_n}{T_{n,max}^{sc}}$. The constraint (\ref{constra:max_T}) (i.e., ${\color{red}\tau}_n+t_n^c \leq \mathcal{T}$) in $\mathbb{P}_4(\boldsymbol{P}, \boldsymbol{X})$ can be written as $r_n\geq \frac{D_n}{\mathcal{T}-t_n^c}$. 
Hence, we can combine the two constraints by denoting $\max\{\frac{\rho\mathcal{C}_n}{T_{n,max}^{sc}},\frac{D_n}{\mathcal{T}-t_n^c}\}$ as $r_n^{min}$.
For easier comprehension and to align with the meaning of $\mathbb{P}_4(\boldsymbol{P}, \boldsymbol{X})$, the constraints (\ref{constra:T_secom_limit}) and (\ref{constra:max_T}) can be combined and written in the following form:
$$r_n\geq r_n^{min}.$$

Hence, Problem $\mathbb{P}_4(\boldsymbol{P}, \boldsymbol{X})$ is transformed into
\begin{subequations} \label{problem:SP2_relax_version}
    \begin{align}
        \min_{\boldsymbol{P}, \boldsymbol{X}} &~\kappa_1\sum_{n\in\mathcal{N}} \frac{(\sum_{k\in\mathcal{K}}p_{n,k})\cdot(D_n+\rho \mathcal{C}_n)}{r_n}, \tag{\ref{problem:SP2_relax_version}}\\
        \text{Subject to,} &~(\ref{constra:sum_pnk_max}), (\ref{constra:sum_xnk}),\notag \\
        & p_{n,k} \leq (x_{n,k})^q P_{n}^{max},~\forall n \in \mathcal{N},~\forall k \in \mathcal{K}, \label{constra:pmax_tighter}\\
        & \sum_{n\in \mathcal{N}}
        \sum_{k\in \mathcal{K}} x_{n,k}\cdot(1-x_{n,k}) \leq 0, \label{constra:xnk_0_1}\\
        & r_n\geq r_n^{min}, \label{constra:rn_min} \\
        & 0 \leq x_{n,k} \le 1,~\forall n \in \mathcal{N},~\forall k \in \mathcal{K}, \label{constra:continuous_xnk}
    \end{align}
\end{subequations}
Despite the transformations we have made, Problem (\ref{problem:SP2_relax_version}) remains challenging to solve. The difficulty of this problem lies in the fact that the objective function is a sum-of-ratios problem, which is NP-complete. Each term of the sum has a concave denominator and a convex numerator. Moreover, constraints (\ref{constra:pmax_tighter}) and (\ref{constra:xnk_0_1}) are non-convex, which further adds to the complexity of the problem.

To better transform the original $\mathbb{P}_4(\boldsymbol{P}, \boldsymbol{X})$, we first address the non-convex constraints (\ref{constra:pmax_tighter}) and (\ref{constra:xnk_0_1}). In the constraint (\ref{constra:pmax_tighter}), the non-convex term $(x_{n,k})^q$ can be approximated by Taylor expansion with a simpler, differentiable function that can be more easily optimized.
Following the Taylor expansion, we have the inequality below:
$$
    (x_{n,k})^q \geq (x_{n,k}^{(i)})^q + q\cdot(x_{n,k}^{(i)})^{q-1}(x_{n,k}-x_{n,k}^{(i)})
$$
Hence, we can replace the constraint (\ref{constra:pmax_tighter}) by the tighter version:
$$
p_{n,k}\leq [(x_{n,k}^{(i)})^q + q\cdot(x_{n,k}^{(i)})^{q-1}(x_{n,k}-x_{n,k}^{(i)})]P_{n}^{max}.
$$

For the constraint (\ref{constra:xnk_0_1}), we plan to employ the Successive Convex Approximation (SCA)
method. In brief, this method iteratively approximates the non-convex function with a convex one, solving the convex function and using the solution to update the approximation in the next iteration. Note that the first-order Taylor series can be utilized to approximate the constraint, which is given by
\begin{align}
   & \sum_{n\in\mathcal{N}}\sum_{k\in\mathcal{K}} x_{n,k}(1-x_{n,k}) \notag \\
   % &\approx 
   & \leq
   \sum_{n\in\mathcal{N}}\sum_{k\in\mathcal{K}} (1-2x_{n,k}^{(i)})(x_{n,k}-x_{n,k}^{(i)})+x_{n,k}^{(i)}(1-x_{n,k}^{(i)}),
\end{align}
where $i$ represents the $i$-th iteration. To simplify the notation, we denote
\begin{align}
    J(\boldsymbol{X}) := \sum_{n\in\mathcal{N}}\sum_{k\in\mathcal{K}} (2x_{n,k}^{(i)}-1)(x_{n,k}-x_{n,k}^{(i)})+x_{n,k}^{(i)}(x_{n,k}^{(i)}-1).
\end{align}
This approximation can be incorporated into the objective function as a penalty term by multiplying a penalty factor $\varsigma$, and $\mathbb{P}_4(\boldsymbol{P}, \boldsymbol{X})$ can be iteratively approximated by the following problem
\begin{subequations} \label{problem:SP2_transformation2}
    \begin{align}
        \min_{\boldsymbol{P}, \boldsymbol{X}} &~\kappa_1\sum_{n\in\mathcal{N}} \frac{(\sum_{k\in\mathcal{K}}p_{n,k})\cdot(D_n+\rho \mathcal{C}_n)}{r_n}-\varsigma \cdot J(x_{n,k}), \tag{\ref{problem:SP2_transformation2}}\\
        \text{Subject to,} &~(\ref{constra:sum_pnk_max}), (\ref{constra:sum_xnk}), 
        (\ref{constra:rn_min}), (\ref{constra:continuous_xnk}), \notag\\
         & \hspace{-40pt} p_{n,k}\leq [(x_{n,k}^{(i)})^q + q\cdot(x_{n,k}^{(i)})^{q-1}(x_{n,k}-x_{n,k}^{(i)})]P_{n}^{max}. \label{constra:pmax_tighter_taylor_series}
    \end{align}
\end{subequations}

Until now, the problem is still non-convex due to the fractional summation form $\sum_{n\in\mathcal{N}} \frac{(\sum_{k\in\mathcal{K}}p_{n,k})\cdot(D_n+\rho \mathcal{C}_n)}{r_n}$. 
Hence, we first transform the fractional form into an epigraph form by adding one auxiliary variable $\boldsymbol{\sigma}:=(\sigma_1, \sigma_2,\dots,\sigma_N)$. Let $\frac{(\sum_{k\in\mathcal{K}}p_{n,k})\cdot(D_n+\rho \mathcal{C}_n)}{r_n} \leq \sigma_n$, and the problem (\ref{problem:SP2_transformation2}) equivalently becomes
\begin{subequations} \label{problem:SP2_transformation3}
\begin{align}
    \min_{\boldsymbol{P}, \boldsymbol{X}, \boldsymbol{\sigma}} & \kappa_1\sum_{n=1}^N \sigma_n -\varsigma \cdot J(\boldsymbol{X}),  
    \tag{\ref{problem:SP2_transformation3}}\\ 
    \text{Subject to,} &~(\ref{constra:sum_pnk_max}), (\ref{constra:sum_xnk}), (\ref{constra:rn_min}), (\ref{constra:continuous_xnk}),
    (\ref{constra:pmax_tighter_taylor_series}),\notag \\
    & \frac{(\sum_{k\in\mathcal{K}}p_{n,k})\cdot(D_n+\rho \mathcal{C}_n)}{\sigma_n}-r_n\leq 0, \label{constra:epi_sigma}
    % &  \hspace{-45pt}
\end{align}
\end{subequations}
where $\boldsymbol{\sigma}:=[\sigma_1,\dots,\sigma_N]$.
The objective function of Problem (\ref{problem:SP2_transformation3}) is convex now. 
Nevertheless, we are still one step away; due to $\frac{(\sum_{k\in\mathcal{K}}p_{n,k})\cdot(D_n+\rho \mathcal{C}_n)}{\sigma_n}$, the constraint (\ref{constra:epi_sigma}) is not convex, and the current problem remains non-convex.
Motivated by the method proposed by \cite{zhao2023human}, we can transform $\frac{\sum_{k\in\mathcal{K}}p_{n,k}}{\sigma_n}$ in the constraint (\ref{constra:epi_sigma}) into 
\begin{align} \label{transformation-yn}
    \frac{\sum_{k\in\mathcal{K}}p_{n,k}}{\sigma_n} = (\sum_{k\in\mathcal{K}} p_{n,k})^2 y_n + \frac{1}{4y_n\sigma_n^2},
\end{align}
where $y_n = \frac{1}{2(\sum_{k\in\mathcal{K}}p_{n,k})\sigma_n}$ and $y_n \in \mathbb{R}^{+}$. The general idea of this method is to fix the value of $y_n$ in each iteration, compute the new results, and then recalculate $y_n$ to proceed to the next round of computation. Also, it has been proven in \cite{zhao2023human} that this alternating method can converge to a stationary point. 

We can further simplify the problem by reducing one constraint.
Note that if the solution of $p_{n,k}$ satisfies constraint (\ref{constra:pmax_tighter_taylor_series}), it also satisfies (\ref{constra:sum_pnk_max}). The reason is as follows:
\begin{align}
    \sum_{k\in\mathcal{K}} p_{n,k} &\leq \sum_{k\in\mathcal{K}} [(x_{n,k}^{(i)})^q + q\cdot(x_{n,k}^{(i)})^{q-1}(x_{n,k}-x_{n,k}^{(i)})]P_{n}^{max} \nonumber\\
    & \leq \sum_{k\in\mathcal{K}} (x_{n,k})^q P_{n}^{max} \leq  \sum_{k\in\mathcal{K}}x_{n,k} P_n^{max} \leq P_n^{max}. \nonumber
\end{align}
Hence, to simplify this problem and facilitate subsequent solving, constraints (\ref{constra:sum_pnk_max}) and (\ref{constra:pmax_tighter_taylor_series}) can be reduced to retaining only constraint (\ref{constra:pmax_tighter_taylor_series}).

Hence, the final form of $\mathbb{P}_4(\boldsymbol{P}, \boldsymbol{X})$ is\\
\textbf{Problem $\mathbb{P}_5(\boldsymbol{P},\boldsymbol{X},\boldsymbol{\sigma})$}:
\begin{subequations} \label{problem:SP2_transformation_final}
\begin{align}
    \min_{\boldsymbol{P}, \boldsymbol{X}, \boldsymbol{\sigma}} & \kappa_1\sum_{n=1}^N \sigma_n -\varsigma \cdot J(\boldsymbol{X}),  
    \tag{\ref{problem:SP2_transformation_final}}\\ 
    \text{Subject to,} &~(\ref{constra:sum_pnk_max}), (\ref{constra:sum_xnk}), (\ref{constra:rn_min}), (\ref{constra:continuous_xnk}), (\ref{constra:pmax_tighter_taylor_series}), \notag \\
    & \hspace{-50pt}\Big((\sum_{k\in\mathcal{K}} p_{n,k})^2 y_n + \frac{1}{4y_n\sigma_n^2}\Big)\cdot(D_n+\rho \mathcal{C}_n) - r_n \leq 0. \label{constra:epi_sigma_yn}
    % &  \hspace{-45pt}
\end{align}
\end{subequations}
When $y_n$ is fixed, Problem $\mathbb{P}_5(\boldsymbol{P},\boldsymbol{X},\boldsymbol{\sigma})$ is convex.
To solve this problem, we adopt KKT conditions, which are sufficient and necessary for achieving the optimal solution.

Then, the partial Lagrange function of Problem (\ref{problem:SP2_transformation_final}) is given below:
\begin{align}
   & L_2(\boldsymbol{X},\boldsymbol{P}, \boldsymbol{\sigma},  \boldsymbol{\beta_k}, \boldsymbol{\lambda}, \boldsymbol{\iota}, \boldsymbol{\nu}) = \kappa_1\sum_{n\in\mathcal{N}}\sigma_n 
   -\varsigma\cdot J(\boldsymbol{X}) \notag \\
   &
   % +\sum_{n\in\mathcal{N}}\varrho_n\cdot(p_n-P_{n}^{max}) 
   + \sum_{k\in\mathcal{K}}\beta_k\cdot(\sum_{n\in\mathcal{N}}x_{n,k}-1)
   + \sum_{n\in\mathcal{N}} \lambda_n\cdot\big( r_{n}^{min} -r_n\big) 
   \notag\\
   &+ \hspace{-2pt}\sum_{n\in\mathcal{N}}\hspace{-2pt}\sum_{k\in\mathcal{K}} \iota_{n,k}\hspace{-1pt}\cdot \hspace{-1pt}[p_{n,k}\hspace{-2pt}-\hspace{-2pt}\big((x_{n,k}^{(i)})^q  q(x_{n,k}^{(i)})^{q-1}(x_{n,k}\hspace{-2pt}-\hspace{-2pt} 
    x_{n,k}^{(i)})\big)P_{n}^{max}]\notag\\
   &    + \sum_{n\in\mathcal{N}}\nu_n\cdot \Big[\Big((\sum_{k\in\mathcal{K}} p_{n,k})^2 y_n  
   +\frac{1}{4y_n\sigma_n^2}\Big)(D_n+\rho \mathcal{C}_n) - r_n\Big],
\end{align}
where 
% $
% \boldsymbol{\varrho}:=[\varrho_1, \dots, \varrho_N]$
$\boldsymbol{\beta_k}:=[\beta_1,\dots, \beta_N]$, $\boldsymbol{\lambda}:=[\lambda_1,\dots,\lambda_N]$, $\boldsymbol{\iota}:=[\iota_{n,k}]_{N\times K}$ ($n\in\mathcal{N},k\in\mathcal{K}$), and $\boldsymbol{\nu}:=[\nu_1,\dots,\nu_N]$ are \mbox{non-negative} Lagrange multipliers. According to KKT conditions, we obtain\\
\textbf{Stationarity}:\\
\begin{align}
\frac{\partial L_2}{\partial x_{n,k}} &= -(\lambda_n+\nu_n)\bar{B}\log_2(1+\frac{p_{n,k}g_{n,k}}{N_0\bar{B}})-\varsigma\cdot(2x_{n,k}^{(i)}-1)\notag\\& +\beta_k-\iota_{n,k} q(x_{n,k}^{(i)})^{(q-1)}P_{n}^{max}=0, \label{kkt:partial_xnk}\\
\frac{\partial L_2}{\partial p_{n,k}} &=-(\lambda_n+\nu_n)\frac{x_{n,k}g_{n,k}}{(1+\frac{p_{n,k}g_{n,k}}{N_0\bar{B}})N_0\ln{2}}
 +\iota_{n,k} \notag\\ & +2\nu_n (D_n+\rho \mathcal{C}_n) y_n (\sum_{k\in\mathcal{K}}p_{n,k}) = 0, \label{kkt:partial_pnk}\\
   % \frac{\partial L_2}{\partial \rho} &= \mathcal{C}\!\sum_{n\in\mathcal{N}}\!( \kappa_1\tau_n(\sum_{k\in\mathcal{K}}x_{n,k}^2c_{n,k} \!+\! \frac{p_{n,k}^2}{4c_{n,k}}) \!+\! \psi_n)  -\kappa_3\frac{\partial A_n}{\partial \rho} =  0,\\
\frac{\partial L_2}{\partial \sigma_n} &=\kappa_1 - \frac{(D_n+\rho \mathcal{C}_n)\nu_n}{2y_n\sigma_n^{-3}} = 0, \label{kkt:partial_sigma}
% \varrho_n\cdot & (p_n-P_{n}^{max})=0, \textcolor{blue}{can~be~deleted}
\end{align}
\textbf{Complementary slackness}:\\
\begin{align}
    \lambda_n\cdot & \big( r_{n}^{min} 
    -r_n\big) = 0, \label{kkt:lambda_complementary_slackness}\\
    \iota_{n,k} \cdot &[p_{n,k} \!-\! \big((x_{n,k}^{(i)})^q \!+\! q(x_{n,k}^{(i)})^{(q-1)}(x_{n,k}-x_{n,k}^{(i)})\big)P_{n}^{max}] \!=\! 0, \label{kkt:iota_complementary_slackness} \\
    \beta_k\cdot  ( &\sum_{n\in\mathcal{N}}x_{n,k}-1)=0, \label{kkt:beta_complementary_slackness}\\
    \nu_n \!\cdot\! \Big[\!\Big(&\!(\sum_{k\in\mathcal{K}} p_{n,k})^2 y_n \!+\! \frac{1}{4y_n\sigma_n^2}\Big)\!(D_n+\rho \mathcal{C}_n) \!-\! r_n\!\Big]\!=\!0, \label{kkt:nu_complementary_slackness}
    \end{align}
    \textbf{Primal feasibility}:\\
    \begin{align}
    \text{(\ref{constra:sum_pnk_max}), (\ref{constra:sum_xnk}), (\ref{constra:rn_min}), (\ref{constra:continuous_xnk}), (\ref{constra:pmax_tighter_taylor_series}), (\ref{constra:epi_sigma_yn})}.\notag
    \end{align}
    \textbf{Dual feasibility}:\\
    \begin{align}    
\lambda_n,&~\iota_{n,k},~\beta_k, \nu_n\geq 0,~\forall~n \!\in\! \mathcal{N},~k\!\in\! \mathcal{K}.
\end{align}

% Through analysis, $\frac{\partial L_2}{\partial x_{n,k}}$ is an equation with $x_{n,k}$, $p_{n,k}$, $\rho$, $\beta_k$, $\nu$, and $\lambda_n$ as variables.
% $\frac{\partial L_2}{\partial p_{n,k}}$ is also an equation with $x_{n,k}$, $p_{n,k}$, $\rho$, $\iota_{n,k}$ and $\lambda_n$ as variables. 
% % Besides, $\frac{\partial L_2}{\partial \rho}$ is an equation with $x_{n,k}$, $p_{n,k}$ and $\psi_n$ as variables.
% Hence, we define 
% \begin{align}
% & \widehat{p}_{n,k}(x_{n,k},p_{n,k},\beta_k,\nu,\lambda_n):= \frac{\partial L_2}{\partial x_{n,k}},\\
% & \Psi_{n,k}(x_{n,k},p_{n,k},\iota_{n,k},\lambda_n):= \frac{\partial L_2}{\partial p_{n,k}}.
% \end{align}

Through analysis, it is easy to obtain from (\ref{kkt:partial_xnk}):
\begin{align}
    p_{n,k} = \widehat{p}_{n,k}(\beta_k,\iota_{n,k},\lambda_n,\nu_n), ~\forall~n \!\in\! \mathcal{N},~k\!\in\! \mathcal{K}, \label{p_nk_lagrange_mul}
\end{align}
where
\begin{align}
    &\widehat{p}_{n,k}(\beta_k,\iota_{n,k},\lambda_n,\nu_n)\notag\\
    &:= [2^{\frac{-\varsigma(2x_{n,k}^{(i)}-1)+\beta_k - \iota_{n,k}q(x_{n,k}^{(i)})^{q-1}P_{n}^{max}}{(\lambda_n+\nu_n)\bar{B}}}-1]\frac{N_0\bar{B}}{g_{n,k}}. \label{p_nk_lagrange_multiplier}
\end{align}
For ease of representation, we define the above Eq. (\ref{p_nk_lagrange_multiplier}) as $\widehat{p}_{n,k}$, representing $p_{n,k}$ as a function of the four Lagrange multipliers $\beta_k,\iota_{n,k},\lambda_n$, and $\nu_n$.

Similarly, from (\ref{kkt:partial_pnk}), we can derive that 
\begin{align} \label{x_nk_lagrange_mul_origin}
     \frac{x_{n,k}\cdot(\iota_{n,k}+2\nu_n y_nD_n\sum_{k\in\mathcal{K}}p_{n,k})(1+\frac{p_{n,k}g_{n,k}}{N_0\bar{B}})N_0\ln{2}}{\lambda_n+\nu_n}.
\end{align}
Moreover, given $p_{n,k} = \widehat{p}_{n,k}(\beta_k,\iota_{n,k},\lambda_n,\nu_n)$, we could also represent $x_{n,k}$ by using the four Lagrange multipliers $\beta_k,\iota_{n,k},\lambda_n$, and $\nu_n$. Hence, given (\ref{p_nk_lagrange_mul}), (\ref{p_nk_lagrange_multiplier}) and (\ref{x_nk_lagrange_mul_origin}), we could define
\begin{equation} \label{x_nk_lagrange_multiplier}
    x_{nk}=\widehat{x}_{n,k}(\beta_k,\iota_{n,k},\lambda_n,\nu_n),~\forall~n \!\in\! \mathcal{N},~k\!\in\! \mathcal{K}.
\end{equation}

Additionally, from (\ref{kkt:partial_sigma}), it could be derived that
\begin{align} \label{sigma_calculation}
    \sigma_n = \sqrt[3]{\frac{\nu_nD_n}{2y_n\kappa_1}}.
\end{align}
Therefore, $\sigma$ can be viewed as a function of the Lagrange multiplier $\nu_n$, and we thereby denote
\begin{equation} \label{sigma_nu}
    \sigma_n = \widehat{\sigma}_n(\nu_n),~\forall~n\in\mathcal{N}.
\end{equation}
Since $\sigma_n>0$, it could be concluded that the Lagrange multiplier $\nu_n>0$ and cannot be $0$.
% Next, we discuss how to calculate $x_{n,k}, p_{n,k}$ under different conditions of Lagrange multipliers $\beta_k$, $\iota_{n,k}$, $\lambda_n$, and  $\nu_n$. 
Thus, according to (\ref{kkt:nu_complementary_slackness}), \begin{align}
 r_n  \!=\!  \!\Big(&\!(\sum_{k\in\mathcal{K}} p_{n,k})^2 y_n \!+\! \frac{1}{4y_n\sigma_n^2}\Big)\!D_n\Big. \label{kkt:nu_complementary_slackness2}
\end{align}

We can only deduce that $\nu_n>0$ based on the existing conditions, but whether the other Lagrange multipliers are $0$ or not remains indeterminate at this point. Consequently, we will proceed with a step-by-step analysis.
% If the obtained $x_{n,k}$ and $p_{n,k}$ satisfy the conditions, the assumption that corresponding Lagrange multipliers are $0$ is not violated. Otherwise, the Lagrange multipliers should be larger than $0$, and other scenarios must be taken into consideration.

\subsection{Discussion of KKT Conditions of Problem $\mathbb{P}_5(\boldsymbol{P}, \boldsymbol{X}, \boldsymbol{\sigma})$} \label{discuss_KKT_SP2}
In this section, we will analyze the values of Lagrange multipliers, $\boldsymbol{\beta}$, $\boldsymbol{\iota}$, $\boldsymbol{\lambda}$ and $\boldsymbol{\nu}$.
To better analyze the problem, we have defined several functions in the previous section, namely $\widehat{p}_{n,k}(\beta_k,\iota_{n,k},\lambda_n,\nu_n)$,
$\widehat{x}_{n,k}(\beta_k,\iota_{n,k},\lambda_n,\nu_n)$
and $\widehat{\sigma}_n(\nu_n)$,
to represent $p_{n,k}$, $x_{n,k}$ and $\sigma_n$ ($n\in\mathcal{N}$,$k\in\mathcal{K}$), respectively.

Also, to simplify the notations, we also define 
\begin{align} \label{x_nk_cap}
    & \overline{x}_{n,k}(\beta_k,\iota_{n,k},\lambda_n,\nu_n) \notag \\
    & = \min\Big(\max\big(\widehat{x}_{n,k}(\beta_k,\iota_{n,k},\lambda_n,\nu_n), 0 \big), 1 \Big)
\end{align}
to make sure that the range of $x_{n,k}$ is between $0$ and $1$.

Moreover, for the convenience of the reader, we have organized the constraints corresponding to each Lagrange multiplier as follows:
\begin{table}[h!]
\centering
\begin{tabular}{|c|c|}
\hline
 & \textbf{Constraints}                                                                                                   \\ \hline
$\beta_k$                     & (\ref{constra:sum_xnk}): $\sum_{n\in\mathcal{N}}x_{n,k}-1 \leq 0$                                                                               \\ \hline
$\iota_{n,k}$                 & (\ref{constra:sum_pnk_max}): $p_{n,k}\!\leq\!\big(\!(x_{n,k}^{(i)})^q \!+\! q(x_{n,k}^{(i)})^{q-1}\!(x_{n,k}\!-\!x_{n,k}^{(i)})\!\big)P_{n}^{max}$              \\ \hline
$\lambda_n$                   & (\ref{constra:rn_min}): $r_n \geq r_{n}^{min}$                                                                                                  \\ \hline
$\nu_n$                       & (\ref{constra:epi_sigma_yn}): $\Big(\!(\sum_{k\in\mathcal{K}} p_{n,k})^2 y_n \!+\! \frac{1}{4y_n\sigma_n^2}\Big)\!(D_n+\rho \mathcal{C}_n)- r_n\!\leq\!0$ \\ \hline
\end{tabular}
\end{table}

% Note that if we discuss all the possible cases thoroughly, it would be tedious, so we group some similar cases together for discussion. First of all, we define ``$\blacksquare$'' as a set to contain the constraints satisfying the corresponding condition 
%     \begin{equation} 
%         \blacksquare  = \left\{
%             \begin{aligned}
%                 & p_n = P_n^{max},~\text{if}~\varrho_n > 0,\\
%                 & r_n = r_n^{min},~\text{if}~\lambda_n>0,\\
%                 & p_{n,k}=p_{n,k} \!-\! \big((x_{n,k}^{(i)})^q \!+\! q(x_{n,k}^{(i)})^{(q-1)}(x_{n,k}-x_{n,k}^{(i)})\big)P_{n}^{max},\notag\\~&\text{if}~\iota_{n,k}>0,\\
%                 & \sum_{n=1}^N x_{n,k} = 1,~\text{if}~\beta_k>0. 
%             \end{aligned}
%         \right. \notag
%     \end{equation}  

    \textbf{Step 1}: First, we could analyze the Lagrange multiplier $\boldsymbol{\beta}$. We first calculate $\widehat{x}_{n,k}(\iota_{n,k}, \lambda_n,\nu_n|\beta_k=0)$ by setting $\beta_k=0$.
    Hence, there are two cases:
        \begin{enumerate}
            \item If $\sum_{n=1}^N \overline{x}_{n,k}(\iota_{n,k},\lambda_n,\nu_n|\beta_k=0) \leq 1$, $\beta_k=0$ holds true.
            \item Otherwise, we cannot set $\beta_k=0$ since it will violate the primal feasibility (\ref{constra:sum_xnk}). Then, $\beta_k>0$ and $\sum_{n=1}^N \overline{x}_{n,k}(\iota_{n,k},\lambda_n,\nu_n|\beta_k>0) = 1$. Given this condition, $\beta_k$ can be represented by other three Lagrange multipliers as $\widehat{\beta_k}(\iota_{n,k}, \lambda_n,\nu_n)$.
        \end{enumerate}
        Given these two cases, it could be obtained that 
        \begin{equation}\label{beta_opt}
            \beta_k^* = \left\{ 
            \begin{aligned}
                &0, \text{if}~\sum_{n=1}^N \overline{x}_{n,k}(\iota_{n,k},\lambda_n,\nu_n|\beta_k=0) \leq 1,\\
               & \widehat{\beta_k}(\iota_{n,k}, \lambda_n,\nu_n),~\text{otherwise}.
            \end{aligned}
            \right.
        \end{equation}

    \textbf{Step 2}: Next, we analyze the Lagrange multiplier $\boldsymbol{\iota}$. Since $\beta_k$ is obtained in Step 1, we will omit it in subsequent steps. By setting $\iota_{n,k}=0$, we could compute $\widehat{p}_{n,k}(\lambda_n,\nu_n|\iota_{n,k}=0)$.
    Similarly, there will be two cases:
    \begin{enumerate}
        \item If $\widehat{p}_{n,k}(\lambda_n,\nu_n|\iota_{n,k}\!=\!0)\leq\bigg((x_{n,k}^{(i)})^q \!+\! q(x_{n,k}^{(i)})^{q-1}\!\Big(\overline{x}_{n,k}(\lambda_n,\nu_n|\iota_{n,k}\!=\!0)\!-\!x_{n,k}^{(i)}\Big)\bigg)P_n^{max}$, $\iota_{n,k}$ can be set to $0$.
        \item Otherwise, $\iota_{n,k}=0$ will result in (\ref{constra:sum_pnk_max}) not being satisfied. Hence, $\iota_{n,k}>0$, and we could obtain $p_{n,k}(\lambda_n,\nu_n|\iota_{n,k}>0)=\Big((x_{n,k}^{(i)})^q \!+\! q(x_{n,k}^{(i)})^{q-1}\!(\overline{x}_{n,k}(\lambda_n,\nu_n|\iota_{n,k}>0)\!-\!x_{n,k}^{(i)})\Big)P_n^{max}$. Then, $\iota_{n,k}$ can be represented by other two Lagrange multipliers $\lambda_n$ and $\nu_n$, and we denote it as $\widehat{\iota}_{n,k}(\lambda_n, \nu_n)$.
    \end{enumerate}
    Here, we could summarize the value of Lagrange multiplier $\boldsymbol{\iota}$ is
    \begin{align} \label{iota_opt}
       \iota_{n,k}^* = \left\{  \begin{aligned}
            & \underline{0,}~
            \text{if}~\widehat{p}_{n,k}(\lambda_n,\nu_n|\iota_{n,k}=0)\leq\bigg((x_{n,k}^{(i)})^q \!+\! q(x_{n,k}^{(i)})^{q-1} \\
            &\times\!\Big(\overline{x}_{n,k}(\lambda_n,\nu_n|\iota_{n,k}=0)\!-\!x_{n,k}^{(i)}\Big)\bigg)P_n^{max}, \\
            & \underline{\widehat{\iota}_{n,k}(\lambda_n,\!\nu_n),}
               ~ \text{otherwise}.
       \end{aligned}
       \right.
    \end{align}
    
    % We denote the Lagrange multipliers equal to $0$ as [$\diamondsuit$, $\Box$, $\vartriangle$].

    \textbf{Step 3}: Until now, we have analyzed the values of $\boldsymbol{\beta}$ and $\boldsymbol{\iota}$. In this step, we analyze the value of Lagrange multiplier $\boldsymbol{\lambda}$. Note that it corresponds to the constraint (\ref{constra:rn_min}). Similarly, we still set $\lambda_n=0$. We then compute $$\widehat{r}_n(\nu_n|\lambda_n=0) = r_n\Big(p_{n,k}(\nu_n|\lambda_n=0), \overline{x}_{n,k}(\nu_n|\lambda_n=0)\Big).$$
    It could still be divided into two cases to discuss:
    \begin{enumerate}
        \item If $\widehat{r}_n(\nu_n|\lambda_n=0)\geq r_n^{min}$, $\lambda_n=0$ is true.
        \item If $\widehat{r}_n(\nu_n|\lambda_n=0)< r_n^{min}$, we should set $\lambda_n>0$ to make sure $\widehat{r}_n(\nu_n|\lambda_n>0)= r_n^{min}$ at least. Thus, given this condition, $\lambda_n$ could be denoted by a function of $\nu_n$: $\widehat{\lambda}_n(\nu_n)$.
    \end{enumerate}

    Therefore, the value of $\boldsymbol{\lambda}$ is summarized as follows:
    \begin{equation} \label{lambda_opt}
        \lambda_n^*=\left\{
        \begin{aligned}
            & 0, ~\text{if}~\widehat{r}_n(\nu_n|\lambda_n=0)\geq r_n^{min},\\
            & \widehat{\lambda}_n(\nu_n), ~\text{if}~\widehat{r}_n(\nu_n|\lambda_n=0)< r_n^{min}.
        \end{aligned}
        \right.
    \end{equation}

    \textbf{Step 4}: Now, we only have one Lagrange multiplier $\boldsymbol{\nu}$ to analyze. 
    In addition, as shown in Eq. (\ref{sigma_calculation}), $\nu_n>0$. 
    Hence, we have
    \begin{align} \label{cal_nu_n_only}
        \Big(\!\sum_{k\in\mathcal{K}} \widehat{p}_{n,k}(\nu_n)^2 y_n \!+\! \frac{1}{4y_n\widehat{\sigma}(\nu_n)_n^2}\Big)\!(D_n+\rho \mathcal{C}_n)- \widehat{r}_n(\nu_n)\!=\!0. 
    \end{align}
    We could find the solution $\nu_n^*$ according to the above equation. Given the optimal $\nu_n^*$, we could use it to calculate $\lambda_n^*$ (i.e., (\ref{lambda_opt})) in Step 3. Then, substitute $(\nu_n^*, \lambda_n^*)$ into (\ref{iota_opt}) in Step 2 to get the optimal $\iota_{n,k}^*$. Finally, given $(\nu_n^*, \lambda_n^*, \iota_{n,k}^*)$, we could use (\ref{beta_opt}) to calculate the optimal $\beta_k^*$.

    Then, from Steps 1--4, we could compute the value of each Lagrange multiplier. The optimal solution $(\boldsymbol{P}^*, \boldsymbol{X}^*, \boldsymbol{\sigma}^*)$ can be obtained by $\widehat{p}_{n,k}(\beta_k,\iota_{n,k},\lambda_n,\nu_n)$ in (\ref{p_nk_lagrange_multiplier}), $\overline{x}_{n,k}(\beta_k,\iota_{n,k},\lambda_n,\nu_n)$ in (\ref{x_nk_cap}) and $\widehat{\sigma}(\nu_n)$ in (\ref{sigma_nu}), respectively.

\begin{theorem} \label{theorem:cal-p-x}
    The optimal solution $(\boldsymbol{P}^*, \boldsymbol{X}^*, \boldsymbol{\sigma}^*)$ of Problem $\mathbb{P}_5$ can be computed by
    \begin{align}
        & p_{n,k}^* = \widehat{p}_{n,k}(\beta_k,\iota_{n,k},\lambda_n,\nu_n) (\text{i.e., (\ref{p_nk_lagrange_multiplier})}) \notag \\
        & x_{n,k}^* = \overline{x}_{n,k}(\beta_k,\iota_{n,k},\lambda_n,\nu_n) (\text{i.e., (\ref{x_nk_cap})}) \notag\\
        & \sigma_n^* = \widehat{\sigma}(\nu_n) (\text{i.e., (\ref{sigma_nu})}) \notag,
    \end{align}
    where $\beta_k$, $\iota_{n,k}$, $\lambda_n$ and $\nu_n$ ($n\in\mathcal{N}, k\in\mathcal{K}$) are obtained from Steps 1--4.
\end{theorem}
    
The above steps illustrate how to compute the optimal solution $(\boldsymbol{P}^*, \boldsymbol{X}^*, \boldsymbol{\sigma}^*)$ by analyzing the values of Lagrange multipliers sequentially.
Alg. \ref{algo:kkt_conditions} streamlines those steps. In Alg. \ref{algo:kkt_conditions}, Line 3 is the first initialization of $\sigma_n$. Line 4 is just a temporary variable for calculating the objective function (\ref{problem:SP2_transformation_final}) in Problem $\mathbb{P}_5$. Besides, Line 9 is to update $y_n$ for each iteration, and $y_n$ is defined in (\ref{transformation-yn}).

\begin{algorithm} 
    \caption{Find the Solution Set of Problem $\mathbb{P}_5(\boldsymbol{P}, \boldsymbol{X}, \boldsymbol{\sigma})$} \label{algo:kkt_conditions}
    % \KwIn{The iteration number $i=0$, $\beta_k=0, \nu_n=0, \iota_{n,k}=0, \lambda_n=0$}
    \SetKwFunction{main}{main}
    \SetKwFunction{CHECKiota}{CHECK\_$\iota_{n,k}$}
    \SetKwFunction{CHECKlambda}{CHECK\_$\lambda_n$}
    \SetKwProg{myalg}{Algorithm}{}{}
    % \myalg{\main{}}{
    
    Given the initial $(\boldsymbol{f}, \rho, \boldsymbol{P}^{(0)}, \boldsymbol{X}^{(0)})$, and the penalty factor $\varsigma$.
    
    Initialize the iteration number $i\leftarrow 0$.

    $\sigma_n^{(0)} = \frac{(\sum_{k\in\mathcal{K}}p_{n,k}^{(0)})\cdot(D_n+\rho \mathcal{C}_n)}{r_n^{(0)}}$, $\forall~n\in\mathcal{N}$.

    $h^{(0)} \leftarrow \kappa_1(\sum_{n\in\mathcal{N}} \sigma_n^{(0)}) - \varsigma\cdot J(\boldsymbol{X}^{(0)})$.
    
    \Repeat{$|h^{i}-h^{i-1}|\leq \epsilon_1$ or the maximum iteration number $I_{max}$ is reached}{
        $i\leftarrow i+1$.

        $y_n^{(i)} \leftarrow \frac{1}{2(\sum_{k\in\mathcal{K}} p_{n,k}^{(i-1)})\sigma_n^{(i-1)}}$, $\forall~n\in\mathcal{N}$.
        
        Find the values of Lagrange multipliers $(\boldsymbol{\nu}, \boldsymbol{\beta}, \boldsymbol{\iota}, \boldsymbol{\lambda})$ through Steps 1--4 in Section \ref{discuss_KKT_SP2}.

        Calculate the solution $(\boldsymbol{P}^{(i)}, \boldsymbol{X}^{(i)}, \boldsymbol{\sigma}^{(i)})$ according to Theorem \ref{theorem:cal-p-x}.
        
        $h^{(i)} \leftarrow \kappa_1(\sum_{n\in\mathcal{N}} \sigma_n^{(i)}) - \varsigma\cdot J(\boldsymbol{X}^{(i)})$.
    
    }
\end{algorithm}

\subsection{Resource Allocation Algorithm}
Next, the resource allocation algorithm is provided in \mbox{Alg. \ref{algo:resource_allo}}.

First, we will provide the feasible solution set $(\boldsymbol{f}, \rho, \boldsymbol{P}, \boldsymbol{X})$ within the range of each optimization variable. Line 1 is to calculate the objective function (\ref{problem:relax_version}) in Problem $\mathbb{P}_2$.
Then, given $(\boldsymbol{P}, \boldsymbol{X})$, we first solve subproblem $\mathbb{P}_3(\boldsymbol{f}, \rho, \mathcal{T})$ by calling \mbox{Theorem \ref{theorem:cal-f-rho}}. After finding the optimal $(\boldsymbol{f}, \rho)$ in $\mathbb{P}_3(\boldsymbol{f}, \rho, \mathcal{T})$, Alg. \ref{algo:kkt_conditions} is called to obtain the solution set $(\boldsymbol{P}, \boldsymbol{X})$. Hence, this is an iterative process, and the resource allocation algorithm will stop until it converges.

\begin{algorithm}
    \caption{Resource Allocation Algorithm} \label{algo:resource_allo}
    \KwIn{The iteration number $i\leftarrow 0$, feasible initial solution $(\boldsymbol{f}^{(0)}, \rho^{(0)}, \boldsymbol{P}^{(0)}, \boldsymbol{X}^{(0)})$, the weight parameters ($\kappa_1, \kappa_2, \kappa_3$).}
    \SetKwFunction{FMain}{\underline{Calculate\_p\_x}}
    \SetKwProg{Fn}{Function}{:}{}

    Use the initial $(\boldsymbol{f}^{(0)}, \rho^{(0)}, \boldsymbol{P}^{(0)}, \boldsymbol{X}^{(0)})$ to calculate $$s^{(0)} \leftarrow \kappa_1\sum_{n\in\mathcal{N}} E_n \!+\! \kappa_2 \mathcal{T} \!-\! \kappa_3 \sum_{n\in\mathcal{N}} A_n(\rho^{(i)}).$$
    
    \Repeat{$|s^{(i)}-s^{(i-1)}|\leq\epsilon_2$ or the maximum iteration number $J_{max}$ is reached}{   

    $i \leftarrow i+1$.
    
    \tcc{First, $\mathbb{P}_3(\boldsymbol{f}, \rho, \mathcal{T})$ is to be solved. Note that $\mathcal{T}$ is an auxiliary variable.}
        Given $(\boldsymbol{P}^{(i-1)}, \boldsymbol{X}^{(i-1)})$, calculate $(\boldsymbol{f}^{(i)}, \rho^{(i)})$ by using \mbox{Theorem \ref{theorem:cal-f-rho}}.
        
        \tcc{The calculated $(\boldsymbol{f}^{(i)}, \rho^{(i)})$ will be used in solving $\mathbb{P}_5(\boldsymbol{P}, \boldsymbol{X}, \boldsymbol{\sigma})$. $\boldsymbol{\sigma}$ is also an auxiliary variable.}
        With $(\boldsymbol{f}^{(i)}, \rho^{(i)})$, obtain $(\boldsymbol{P}^{(i)}, \boldsymbol{X}^{(i)})$ by calling Alg. \ref{algo:kkt_conditions}.

        Use $(\boldsymbol{P}^{(i)}, \boldsymbol{f}^{(i)}, \boldsymbol{X}^{(i)}, \rho^{(i)})$ to calculate $$s^{(i)} \leftarrow \kappa_1\sum_{n\in\mathcal{N}} E_n \!+\! \kappa_2 \mathcal{T} \!-\! \kappa_3 \sum_{n\in\mathcal{N}} A_n(\rho^{(i)}).$$
    }
    % \Fn{\FMain{$f_n$}}{
    %     Calculate $$\tau_n = \frac{1}{r_n(x_{n,k}^{(i-1)}, p_{n,k}^{(i-1)})}, \theta_n=\frac{G_n(x_{n,k}^{(i-1)}, p_{n,k}^{(i-1)})}{r_n(x_{n,k}^{(i-1)}, p_{n,k}^{(i-1)})},$$

    %     Obtain $((x_{n,k}^{(i-1)}, p_{n,k}^{(i-1)})$ through calling Algorithm \ref{algo:kkt_conditions}.    

    % }
\end{algorithm}

\subsubsection{Time complexity}
We will analyze the algorithm's time complexity based on the number of users 
$N$ and the number of subcarriers $K$. Also, the loop section is the main part of Alg. \ref{algo:kkt_conditions} and \ref{algo:resource_allo}. Therefore, we will only analyze the time complexity of the loop section.

Since Alg. \ref{algo:kkt_conditions} will be called in Alg. \ref{algo:resource_allo}. We first analyze the time complexity of Alg. \ref{algo:kkt_conditions}. The time complexities of lines 6--7 are $\mathcal{O}(1)$ and $\mathcal{O}(N)$, respectively. Line 8 follows Steps 1--4 in Section \ref{discuss_KKT_SP2}. Besides, there are $N$-dimensional and $N\times K$-dimensional variables. Hence, to find the solution set in Line 8, the time complexity will be $\mathcal{O}(K+NK+2N)$. The time complexities of Lines 9--10 are $\mathcal{O}(2NK+N)$ and $\mathcal{O}(NK)$. In addition, the maximum iteration number is $I_{max}$. In summary, the time complexity of Alg. \ref{algo:kkt_conditions} is $\mathcal{O}((4NK+3N+K)I_{max})$.

In Alg. \ref{algo:resource_allo}, the time complexity of Line 3 is $\mathcal{O}(1)$. In Line 4, the bisection method is utilized in Theorem \ref{theorem:cal-f-rho}. However, the bisection method is independent of $N$ or $K$, and it is related to the precision of the solution. The overall time complexity of Line 4 will be $\mathcal{O}(N)$. 
Line 5 calls Alg. \ref{algo:kkt_conditions}, and its time complexity is $\mathcal{O}((4NK+3N+K)I_{max})$ as we have analyzed in the previous paragraph. The time complexity of Line 6 is $\mathcal{O}(N)$. Since this is a loop process, and the maximum iteration number is $J_{max}$, the time complexity of Alg. \ref{algo:resource_allo} will be $\mathcal{O}((2N+(4NK+3N+K)I_{max})J_{max})$.

\subsubsection{Convergence analysis}
In subproblem $\mathbb{P}_3(\boldsymbol{f}, \rho, \mathcal{T})$, we can find the optimal $(\boldsymbol{f}, \rho)$. In subproblem $\mathbb{P}_4(\boldsymbol{P}, \boldsymbol{X})$, we transform the original $\mathbb{P}_4(\boldsymbol{P}, \boldsymbol{X})$ into Problem $\mathbb{P}_5(\boldsymbol{P}, \boldsymbol{X}, \boldsymbol{\sigma})$. We can find the optimal $(\boldsymbol{P}, \boldsymbol{X})$ of Problem $\mathbb{P}_5(\boldsymbol{P}, \boldsymbol{X}, \boldsymbol{\sigma})$ at each iteration. Therefore, by iteratively solving $\mathbb{P}_3(\boldsymbol{f}, \rho, \mathcal{T})$ and Problem $\mathbb{P}_5(\boldsymbol{P}, \boldsymbol{X}, \boldsymbol{\sigma})$, the proposed resource allocation algorithm will ultimately converge.

\subsubsection{Fairness analysis}
Our optimization problem is to minimize energy and FL time consumption and maximize accuracy. The resource allocation algorithm is designed to distribute the transmission power, bandwidth (i.e., subcarriers), computation frequency, and compression rate for FedSem across all devices involved in the system.
The optimization framework ensures that no device's total FL time (comprising both computation and transmission times) exceeds a predefined maximum limit, $T_{FL}$ (or $\mathcal{T}$). 
By setting $T_{FL}$ for each device, our resource allocation algorithm promises that the computation frequency, transmission power, and bandwidth are allocated within this limit, thereby preventing resource monopolization by any single device.
This constraint inherently promotes fairness by guaranteeing that the resource allocation does not favor any particular device disproportionately.
Additionally, for the SemCom process, we further guarantee fairness by limiting the SemCom transmission time to $T_{n,max}^{sc}$ and standardizing the compression rate across all devices. This univariate compression rate promises that each device experiences the same level of data compression efficiency, contributing to equitable performance in the SemCom aspect of our system.
Also, we admit that the performance of each device's autoencoder, particularly in terms of accuracy and model-specific metrics, is beyond the scope of this paper. Our primary focus is allocating resources and guaranteeing that system-level constraints are met.

\section{Numerical Results} \label{sec:results}

In this section, we  examine the proposed resource allocation algorithm from the following perspectives:
\begin{enumerate}[{1)}]
    \item We investigate the impact of three weight variables on resource allocation, specifically observing how energy consumption, time consumption, and model accuracy fluctuate with changes in weight.
    \item Additionally, the impact of the maximum transmission power $P_{n}^{max}$ on the proposed resource allocation algorithm is also studied.
    \item Besides, we also investigate the impact of the number of users and subcarriers.
    \item The impact of different SemCom task workloads on our proposed algorithm is also studied.
    \item We examine the relationship between the weight parameter $\kappa_3$, the compression rate $\rho$ and the model accuracy.
    \item Finally, the comparison of the resource allocation algorithm with the approximate exhaustive search is discussed.
\end{enumerate}

For the default parameter setting, the number of devices $N$ is $10$, and the number of subcarriers $K$ is set to $50$. The devices are distributed uniformly within a circular area with a radius of $500$ m. The pass loss model is $128.1 + 37.6 \log(\textit{distance}~\text{in km})$, and this model incorporates a shadow fading component with a standard deviation of $8$ dB. The Gaussian noise's power spectral density, denoted as $N_0$, is measured at $174$ dBm/Hz. Each device is tasked with uploading a data size of $2.81\times 10^4$ bits. In addition, every device contains $500$ samples. The number of CPU cycles for each sample, represented as $c_n$, is randomly selected from a range between $[1, 3] \times 10^4$. The effective switched capacitance is established at $10^{-28}$. For all devices, the maximum frequency, $f^{max}_n$, and the maximum transmission power, $p_n^{max}$, are set to $2$ GHz and $20$ dBm, respectively. Furthermore, we default the number of local iterations, $\eta$, to $10$. The total bandwidth $B$ is $20$ MHz. We also assume that $L=10$, which means there are $10$ rounds of semantic communications for each device. The maximum SemCom transmission time $T_{n,max}^{sc}$ is set to $20$ s. The size $C_{n,l}$ of semantic compressed information is $4.15 \times 10^6$ bits.
Besides, $q$ in the inequality (\ref{constra:pmax_tighter_taylor_series}) is set to $2$. We also summarize the parameter settings in Table \ref{tab:parameter}. 

In addition, the default accuracy function $A_n(\rho)$ utilized in the optimization problem is obtained from experimental results rather than being explicitly defined by an analytical formula, and its derivative $\frac{\partial A_n(\rho)}{\partial \rho}$ is approximated numerically.
To obtain the equation, we evaluate the object detection accuracy (measured by mAP) at several discrete compression rates $\rho$ by running YOLOv5 \cite{yolov5} on the COCO dataset. Then, we fit a concave function utilizing the above points in MATLAB, and we obtain the function $A_n(\rho)=0.6356\cdot\rho^{0.4025}$.

\begin{table*}
\centering
\caption{The default parameter settings.} \label{tab:parameter}
\begin{tabular}{|l|l|}
\hline
The number of devices $N$                                                                & $10$                                                \\ \hline
The number of subcarriers $K$                                                            & $50$                                                \\ \hline
The pass loss model                                                                      & $128.1 + 37.6 \log(\textit{distance}~\text{in km})$ \\ \hline
Gaussian noise’s power spectral density $N_0$                                            & $174$ dBm/Hz                                        \\ \hline
% \begin{tabular}[c]{@{}l@{}}
The number of CPU cycles for each sample $c_n$
% \end{tabular} 
& $[1, 3]\times 10^4$                                 \\ \hline
The number of samples $d_n$ on each device                                               & $500$                                               \\ \hline
The effective switched capacitance $\xi$                                                 & $10^{-28}$                                          \\ \hline
The maximum frequency $f^{max}_n$                                                        & $2$ GHz                                             \\ \hline
The maximum transmission power $p_n^{max}$                                               & $20$ dBm                                            \\ \hline
The number of local iterations $\eta$                                                    & $10$                                                \\ \hline
The total bandwidth $B$                                                                  & $20$ MHz                                            \\ \hline
The maximum SemCom transmission time $T_{n,max}^{sc}$                                    & $20$ s                                              \\ \hline
The size $C_{n,l}$ of semantic compressed information                                    & $4.15 \times 10^6$ bits                             \\ \hline
$q$ in the inequality (\ref{constra:pmax_tighter_taylor_series})                                    & $2$                             \\ \hline
\end{tabular}
\end{table*}

% 10 devices. 5 groups. 2 devices per group.

% load of semantic communication task.

%$x = 4.15 \times 10^6 \times 2/3 $

%  load types:
% heavy: $16x$
% slightly heavy: $8x$
% medium: $4x$
% slightly light: $2x$
% light: $x$

%  load types:
% heavy: $10\times16x$
% slightly heavy: $10\times8x$
% medium: $10\times4x$
% slightly light: $10\times2x$
% light: $10x$

%  load types:
% heavy: $50\times16x$
% slightly heavy: $50\times8x$
% medium: $50\times4x$
% slightly light: $50\times2x$
% light: $50x$

\subsection{The Impact of Weight Parameters}
\begin{figure*}[t!]
    \centering
    \includegraphics[width=0.9\linewidth]{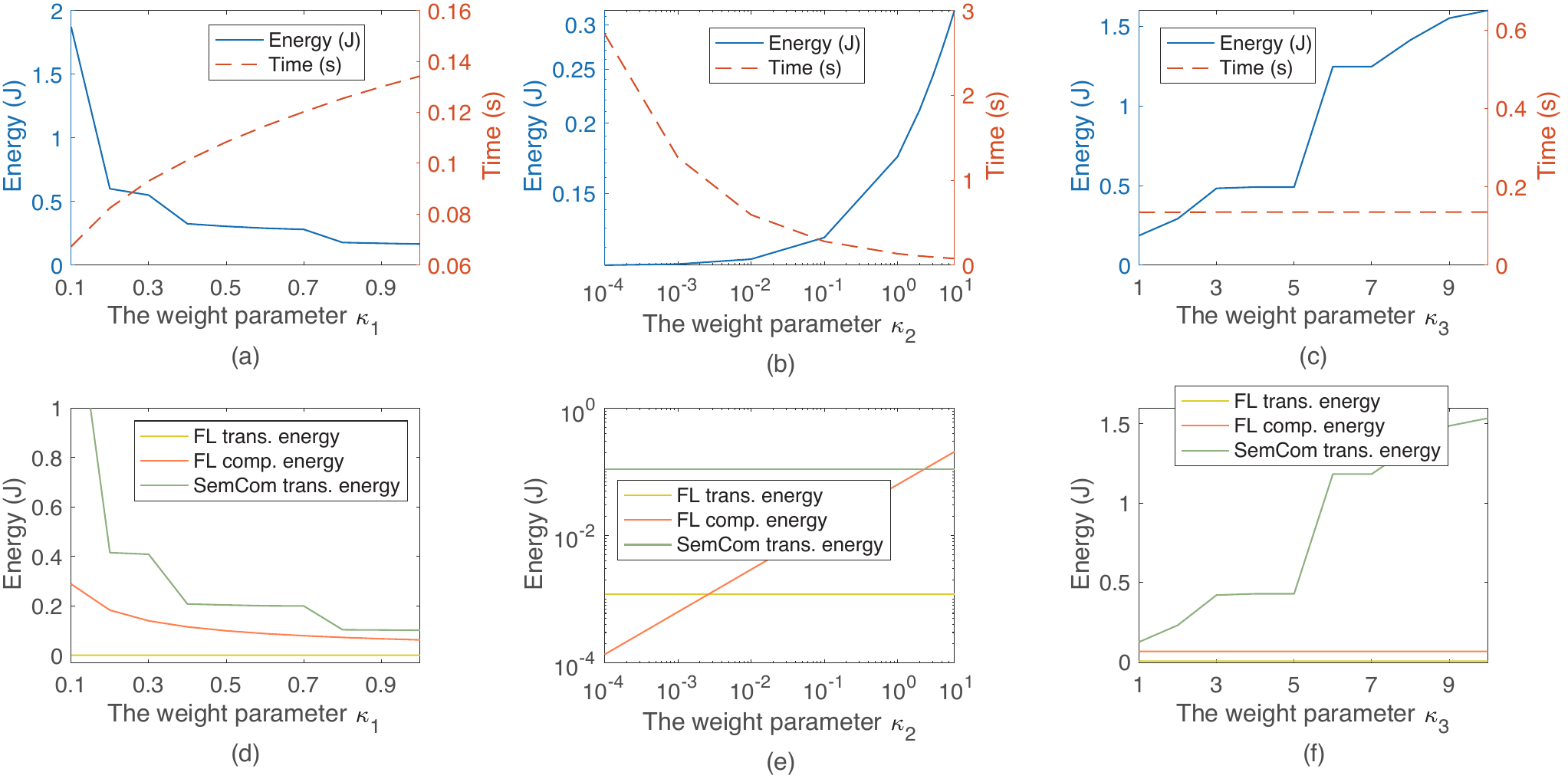}
    \caption{The energy and time consumption under different weight parameters $\kappa_1$, $\kappa_2$ and $\kappa_3$. Note that here when each subfigure has a varying weight parameter, the other weight parameters are set as $1$.}
    \label{fig:e_t_diff_params}
\end{figure*}

In this section, we investigate the impact of the three weight parameters $\kappa_1$, $\kappa_2$ and $\kappa_3$ on the energy and time consumption, as shown in Fig. \ref{fig:e_t_diff_params}.
Figs. \ref{fig:e_t_diff_params}(a)-(c) depict the trends in energy and time consumption as each weight parameter increases.
Besides, Figs. \ref{fig:e_t_diff_params}(d)-(e) illustrate the trend of each energy consumption (i.e., FL transmission energy, FL computation energy, and SemCom transmission energy) and provides insights into which part of the system consumes more energy and how the resource allocation algorithm optimizes this.

Fig. \ref{fig:e_t_diff_params}(a) shows that increasing $\kappa_1$ would place more emphasis on minimizing energy consumption. Accordingly, as the increase of $\kappa_1$, the energy decreases while the FL time consumption increases. Correspondingly, Fig. \ref{fig:e_t_diff_params}(d) reveals that as $\kappa_1$ becomes larger, most components of energy decrease. The trend in FL transmission energy is not obvious. Also, the value of FL transmission energy is much less than that of the other two components, and this may be due to the sufficient bandwidth. This suggests that the resource allocation algorithm prioritizes computation and SemCom energy savings over FL transmission energy.
% Since the trend in FL transmission energy is not obvious, we draw a separate subplot, and the overall trend is also downward.

Besides, Fig. \ref{fig:e_t_diff_params}(b) reveals that higher $\kappa_2$ prefers to reduce the time consumption as the other two weight parameters remain constant. Hence, the increase in energy and decrease in time indicates the algorithm's sensitivity to $\kappa_2$.
Below \mbox{Fig. \ref{fig:e_t_diff_params}(b)}, Fig. \ref{fig:e_t_diff_params}(e) illustrates that the change of $\kappa_2$ has no impact on FL transmission and SemCom transmission energy. 
% This is because the original optimization problem is decomposed into two subproblems, and the optimization of the FL time consumption is divided into $\mathbb{P}_3(\boldsymbol{f}, \rho, \mathcal{T})$ (note that $\mathbb{P}_3(\boldsymbol{f}, \rho, \mathcal{T})$ is to optimization the CPU frequency $f_n$ and the FL time consumption $T_{FL}$), $\kappa_2$ does not take effect in $\mathbb{P}_4(\boldsymbol{P}, \boldsymbol{X})$, in other words,
This is because the FL time optimization does not alter transmission-related parameters. Specifically, in our decomposition approach, the optimization of FL time consumption is handled within $\mathbb{P}_3(\boldsymbol{f}, \rho, \mathcal{T})$, which adjusts CPU frequency $f_n$ but does not impact the transmission power allocation in $\mathbb{P}_4(\boldsymbol{P}, \boldsymbol{X})$.
Consequently, $\kappa_2$ does not contribute to FL transmission energy. For SemCom transmission energy, it is only related to SemCom transmission delay, which is only related to the constraint (\ref{constra:T_secom_limit}) and not to $\kappa_2$.
These observations indicate that while reducing time consumption may improve overall system responsiveness, it comes at the cost of increased computation energy, highlighting the challenge of balancing energy efficiency and latency in resource allocation.

In addition, increasing $\kappa_3$ would prioritize accuracy over energy and time.
Thus, as shown in Fig. \ref{fig:e_t_diff_params}(c), the energy consumption increases along with $\kappa_3$. Nevertheless, the time does not change as $\kappa_3$ changes since the time refers to the FL time consumption, which is not correlated with SemCom in our problem (i.e., it is not correlated with $\rho$, either). 
This suggests that prioritizing accuracy does not necessarily slow down the FL process but instead increases communication costs.
However, because $\kappa_3$ increases, which in turn requires that the compression rate $\rho$ also increases to improve accuracy, there is a greater impact on the transmission energy of SemCom (i.e., more communication resources are allocated to the device), and the energy consumption increases as well.
Accordingly, Fig. \ref{fig:e_t_diff_params}(f) describes the relation between $\kappa_3$ and each energy. It confirms our analysis that an increase in $\kappa_3$ leads to an increase in the compression rate $\rho$ and thus to an increase in the energy consumption of the SemCom transmission. Moreover, $\kappa_3$ has no effect on the other two energy consumption.

These findings highlight the fundamental interplay between energy, time, and accuracy in the resource allocation process. While prioritizing energy reduces power consumption, it increases latency. Conversely, minimizing time consumption requires more energy. Maximizing accuracy, on the other hand, mainly affects transmission energy without significantly affecting computation time.

\subsection{The Impact of $P_n^{max}$}
Apart from weight parameters, we also study the impact of different maximum transmission power $P_{n}^{max}$. Our focus is on two critical metrics: energy consumption and time efficiency. The algorithm's performance is benchmarked against four established baseline methods: Equal Allocation, Communication Optimization Only, Computation Optimization Only, and Random Allocation. 
Here is a detailed description of each baseline method:
\begin{itemize}
    \item \underline{Equal Allocation}: All devices are assigned the same number of subcarriers and also have the same transmission power $p_n$. The computational frequency $f_n$ of each device is set to $1$ GHz. The compression rate $\rho$ is set to $1$.
    \item \underline{Communication Optimization Only}: Only the communication process is optimized, meaning that only the transmission power $\boldsymbol{P}$ and the subcarrier allocation indicator matrix $\boldsymbol{X}$ will be optimized.
    However, the computation process remains unoptimized, with the computational frequency $\boldsymbol{f}$ and the compression rate $\rho$ fixed as constants.
    The computational frequency $\boldsymbol{f}$ is set to a random value between $[0.5, 1.5]$ GHz. The compression rate $\rho$ is set to $1$. 
    \item \underline{Computation Optimization Only}: In this scenario, only the computation procedure is optimized, whilst the communication process is not enhanced. 
    Since the communication process is not optimized, the transmission power $\boldsymbol{P}$ is directly set to its maximum value $P_n^{max}$, and the subcarrier allocation indicator matrix $\boldsymbol{X}$ evenly assigns subcarriers to each device equally.
    \item \underline{Random Allocation}: The subcarrier allocation indicator matrix $\boldsymbol{X}$, the transmission power $\boldsymbol{P}$, and the computational frequency $\boldsymbol{f}$ are selected uniformly at random from the feasible region of our optimization {problem}~$\mathbb{P}_1$. 
    The feasible region can be calculated by using the default parameter settings listed in Table \ref{tab:parameter}. If the random selection does not satisfy the constraints, we will continue searching a feasible solution randomly.
    The compression rate $\rho$ is set to $1$.
\end{itemize}

\begin{figure*}[t!]
    \centering
    \includegraphics[width=1\linewidth]{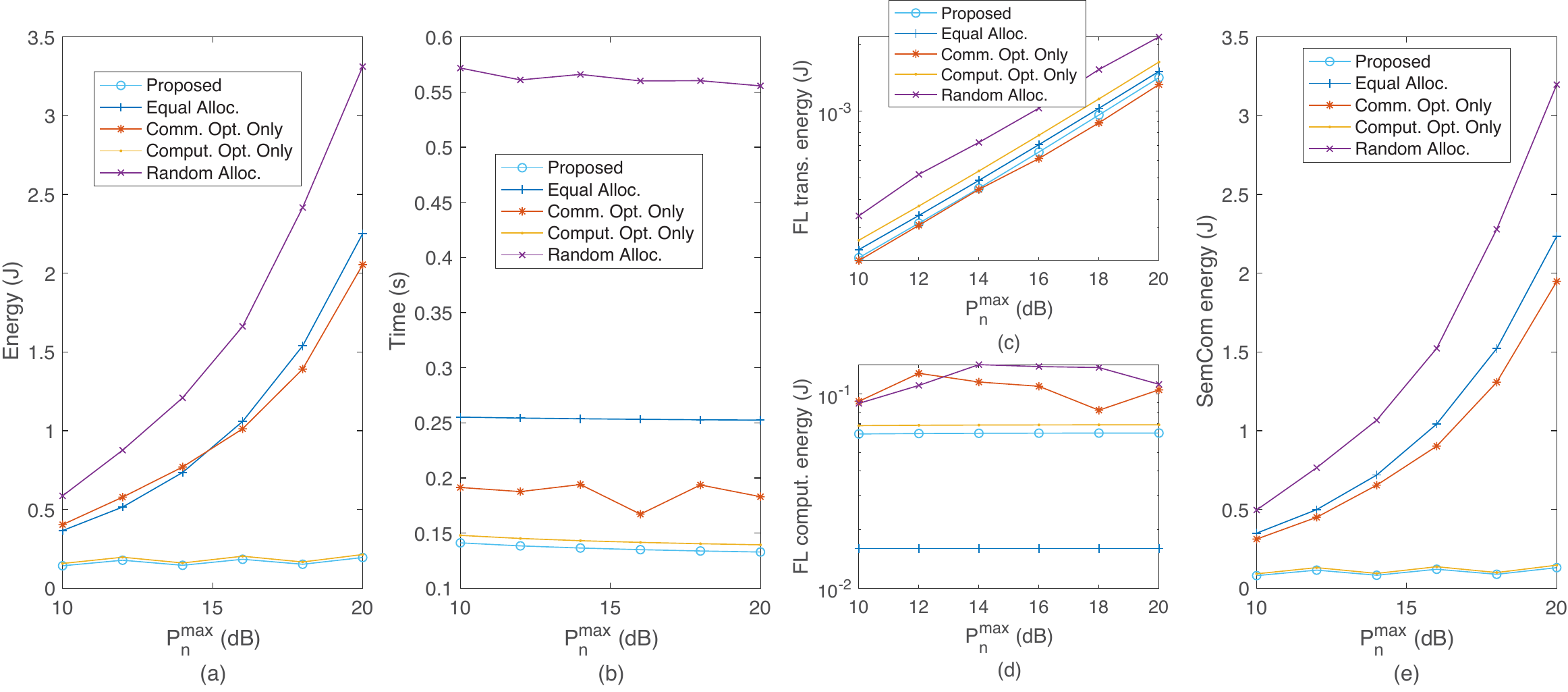}
    \caption{The energy and time consumption under different $P_{n}^{max}$, where $P_{n}^{max}$ is the maximum transmission power. Here we have $\kappa_1=\kappa_2 = \kappa_3=1$. The compression rate $\rho=1$.}
    \label{fig:diff_maximum_transmission_powers}
\end{figure*}
As shown in Fig. \ref{fig:diff_maximum_transmission_powers}, these sub-plots exemplify the effectiveness of our proposed algorithm.

Fig. \ref{fig:diff_maximum_transmission_powers}(a) depicts that the total energy consumption grows with $P_{n}^{max}$ for all schemes. 
This increase occurs because, with higher transmission power budgets, devices can push for faster communication, which in turn raises their transmission energy. Notably, our proposed algorithm still maintains the lowest total energy usage across all maximum transmit power levels. This can be attributed to its joint optimization of communication and computation resources, rather than over-provisioning one resource at the expense of the other. Thus, it achieves a balance that avoids unnecessary energy expenditure on either front.
% Specifically, our algorithm shows an overall significant reduction in energy and time usage compared to the Equal and Random Allocation strategies. This efficiency can be attributed to its adaptive resource management, which minimizes unnecessary transmission and computation costs.
Additionally, the difference in the performance of our proposed algorithm and Computation Optimization Only is less obvious. 
% This may be due to the sufficient bandwidth. The sufficient bandwidth only allows the optimization algorithm to optimize a limited amount of space compared to the computational resources. Thus, Computation Optimization Only performs better than Communication Optimization Only, and is also closer to our proposed algorithm.
This outcome implies the fact that, with sufficient bandwidth, further optimizing communication resources alone may have a diminished marginal benefit compared to optimizing computation. In other words, once the system has already achieved efficient communication (due to ample bandwidth and moderate power), the bottleneck may shift to the computation side, making purely ``Computation Optimization Only'' solutions competitive. Besides, ``Communication Optimization Only'' performs worse than ``Computation Optimization Only'', indicating that ignoring computation‐related overheads leads to suboptimal energy usage.

Figs. \ref{fig:diff_maximum_transmission_powers}(c) and (d) reveal the breakdown of energy consumption and offer insights into the algorithm's internal efficiency. Obviously, for the FL transmission energy, our proposed algorithm and ``Communication Optimization Only'' lead the performance, which demonstrates the effective communication scheduling. However, for the FL computation energy, Equal Allocation is the best compared with other strategies, because it indiscriminately allocates resources in a way that incidentally achieves lower local computation energy than other strategies.
However, “Equal Allocation” fails to account for the total system overhead, especially for transmission costs, leading to higher overall energy than the proposed scheme.
Another energy component, the SemCom transmission energy (Fig. \ref{fig:diff_maximum_transmission_powers}(e)), can also been seen improvements from our joint optimization. 
As $P_{n}^{max}$ increases, devices can transmit semantic information more rapidly, but only our proposed algorithm controls both SemCom energy and FL energy to avoid a disproportionate increase in energy.
% This superiority is because Equal Allocation does not consider the overall optimization, while our proposed algorithm needs to consider other aspects of energy, including FL transmission energy and SemCom energy.

In terms of time efficiency, Fig. \ref{fig:diff_maximum_transmission_powers}(b) exhibits that the proposed algorithm has a balance between time consumption and energy usage.
% Fig. \ref{fig:diff_maximum_transmission_powers}(e) shows a notable performance of the proposed algorithm in terms of SemCom transmission energy, which is crucial for practical applications.

% Fig. \ref{fig:diff_maximum_transmission_powers}(b) highlights the time needed for completing each round of FL. 
While higher $P_n^{max}$ generally reduces communication latency (and thus total time), it may increase transmission energy. Our approach navigates this tradeoff by adapting resource allocations to the system’s current power limits: whenever more power is available, the algorithm strategically shortens communication time without exploiting on energy‐hungry transmissions. This synergy is evident in how our method’s total time remains favorably low across different $P_n^{max}$.

Moreover, it is noticeable that as $P_{n}^{max}$ increases, the energy increases while the time consumption decreases for our proposed algorithm. This phenomenon appears because with the increase of $P_{n}^{max}$, it is more possible to find a better solution to minimize FL time, and thus, the energy increases.

\subsection{The Impact of the Number of Users and the Number of Subcarriers}
In OFDMA, the number of users $N$ and the number of subcarriers $K$ definitely have an influence on the whole system. Hence, we also investigate the influence of $N$ and $K$ on the proposed algorithm.

\begin{figure}[t!]
    \centering
    \includegraphics[width=0.95\linewidth]{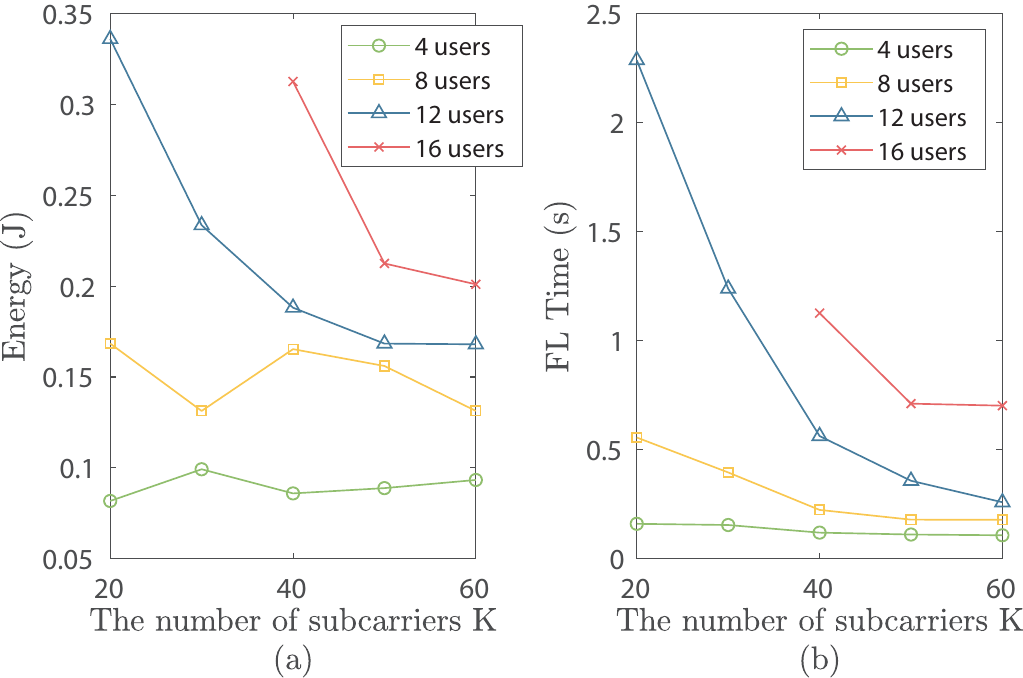}
    \caption{The energy and time consumption under different subcarriers and users. The maximum transmission power $P_{n}^{max}$ is set to $20$ dB. Here we have $\kappa_1=\kappa_2=\kappa_3=1$. The compression rate $\rho = 1$.}
    \label{fig:diff_subcarriers_users}
\end{figure}

Fig. \ref{fig:diff_subcarriers_users}(a) reveals a direct correlation between energy consumption and the number of subcarriers. As the number of subcarriers increases from $20$ to $60$, energy consumption and time consumption show a decreasing trend roughly. This decrease can be caused by the increase in communication resources (i.e., more subcarriers). More communication resources can help the proposed algorithm find more optimal solutions to assign the subcarrier, transmission power and CPU frequency for each user.
Moreover, under the same $K$, as the number of users increases, the energy also increases, which can be attributed to the incremental demand for transmission and computation. The trend also indicates that additional energy is required to manage multiple users. Besides, when the number of users is relatively sparse, for instance, $N=4$ and $K=20$, communication resources are more abundant. As $K$ becomes larger, the changes in energy consumption and delay are insignificant.

Additionally, Fig. \ref{fig:diff_subcarriers_users}(b) shows an overall relation between FL time and the number of subcarriers. It could be observed that the increase of subcarriers leads to an overall decrease in FL time consumption for each user scenario. Also, as the number of users doubles from $4$ to $8$, and then to $16$, the FL time also escalates. This trend could be due to the increase in the number of users and the limited communication resources. 
With limited communication resources, as the number of users increases, the amount of communication resources that can be allocated to each user decreases, resulting in increased latency.

Both Figs. \ref{fig:diff_subcarriers_users}(a) and \ref{fig:diff_subcarriers_users}(b) reveal that the performance of our resource allocation algorithm is sensitive to the number of devices and the number of subcarriers. 
Specifically, when the number of devices or subcarriers is small, the energy consumption and the FL time consumption are also small.
In contrast, a larger number of devices will introduce significant communication overhead, which can increase energy and time consumption.

\subsection{The Impact of the SemCom Task Workloads}
To investigate the impact of different SemCom task workloads on our proposed algorithm, we set the size $\mathcal{C}_n$ of the semantic compressed information differently. Note that for the other settings, we follow the default settings stated in the second paragraph of Section \ref{sec:results}. 
\begin{figure}[t!]
    \centering
    \includegraphics[width=\linewidth]{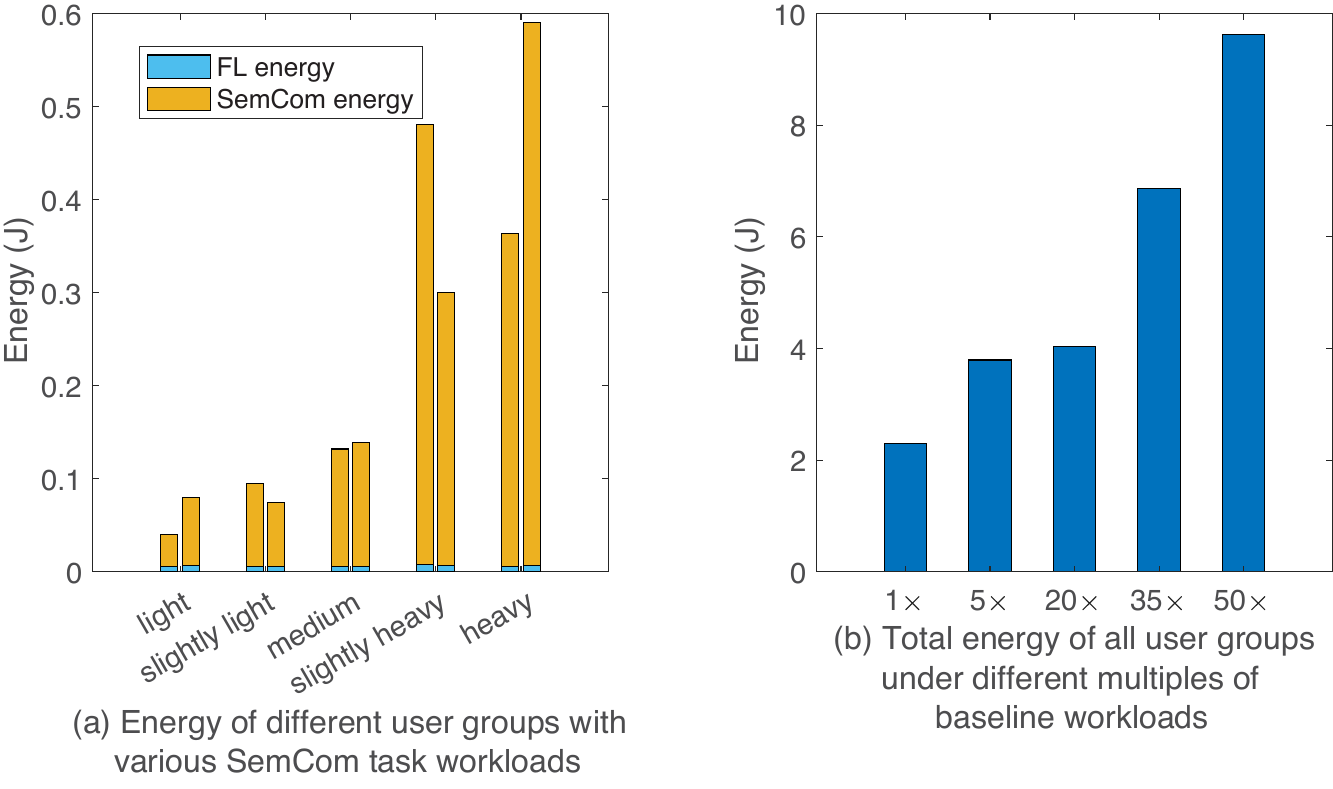}
    \caption{The energy consumption under different SemCom task workloads (i.e., the size $\mathcal{C}_n$ of the semantic compressed information). Here we have $\kappa_1=\kappa_2=\kappa_3=1$.}
    \label{fig:diff-semcom-data}
\end{figure}
Fig. \ref{fig:diff-semcom-data} contains two subfigures. In Fig. \ref{fig:diff-semcom-data}(a), we divide $10$ users into $5$ groups and assign various SemCom task workloads to each group. 
The task workloads are categorized into five types: 
\begin{itemize}
    \item Light: the size of transmitted data size during SemCom is $10^6$ bits. We set this size as the baseline workload $C$, i.e., $C=10^6$.
    \item Slightly light: $2C$.
    \item Medium: $4C$.
    \item Slightly heavy: $8C$.
    \item Heavy: $16C$.
\end{itemize}
It could be observed from Fig. \ref{fig:diff-semcom-data}(a) that the energy increases as the task workloads become more heavy. Since only SemCom task workloads change, there is no apparent variation in the FL energy. The differences in energy within the same group of users are due to their different locations, and thus they cannot consume exactly the same amount of energy.

Besides, we set the task workloads $(C,2C,4C,8C,16C)$ of Fig. \ref{fig:diff-semcom-data}(a) as baseline workloads. Fig. \ref{fig:diff-semcom-data}(b) investigates the changes in total energy of all user groups under different multiples of baseline workloads. 
It is evident that as the workload multiples increase, the total energy consumption also increases.

In conclusion, as shown in Fig. \ref{fig:diff-semcom-data}, the total energy consumption increases with higher workload multiples. This trend is primarily due to the increase in communication overhead that requires more transmission power.
In addition, the performance of our resource allocation algorithm varies under different workload distributions. When workloads are relatively low (e.g., light or slightly light), the algorithm efficiently allocates resources without significant resource contention. However, as the workloads increase to slightly heavy or heavy, resource competition intensifies, which results in higher energy consumption in task execution.

\subsection{Accuracy}
In this section, we investigate how the weight parameter $\kappa_3$ affects the choice of the compression rate $\rho$, and analyze the relation between $\rho$ and the accuracy.
The configuration of the encoder includes a convolution layer (utilizing $5\times 5$ kernel), an encoding sequence (comprising a hyperbolic tangent activation function and a convolution layer), a $2\times 2$ max pooling layer and a second encoding sequence (replicating the first encoding sequence and another hyperbolic tangent activation function). The decoder just mirrors the encoder. For the compression rate $\rho \leq 0.5$, we add one more max pooling layer to reduce the spatial dimensionality. 
% \textcolor{blue}{}

\begin{figure*}[t!]
    \centering
    \includegraphics[width=0.8\linewidth]{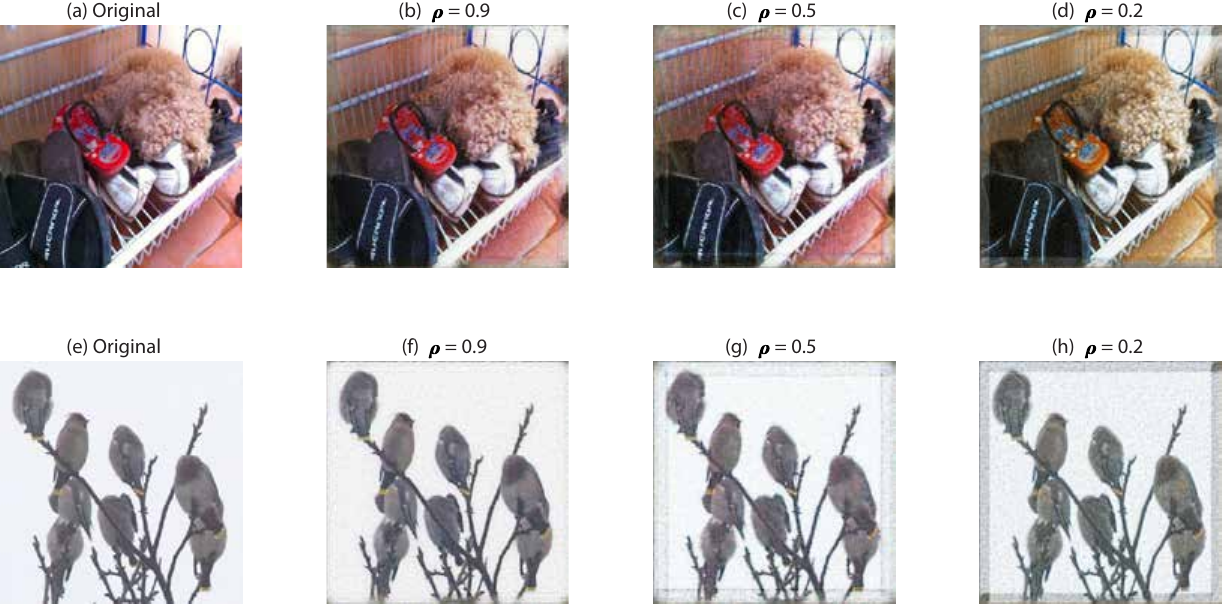}
    \caption{Samples of reconstructed images under different compression rates $\rho$.}
    \label{fig:reconstructed_images}
\end{figure*}

\begin{figure}[t!]
    \centering
    \includegraphics[width=0.95\linewidth]{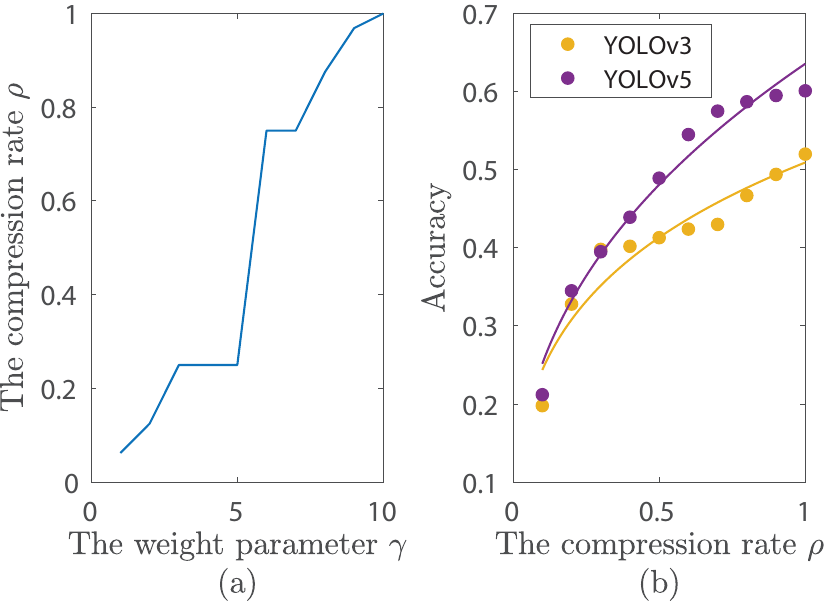}
    \caption{(a) The change of compression rate $\rho$ under different weight parameters $\kappa_3$. Here we have  $\kappa_1=\kappa_2=1$. (b) The accuracy under different compression rates $\rho$.}
    \label{fig:rho_k3_acc}
\end{figure}

Besides, note that the utilized autoencoder primarily serves to assess the impact of our proposed resource allocation algorithm on the overall FedSem system performance. The focus is not on optimizing the model itself or enhancing its accuracy. In this paper, it is critical to observe the changes in system performance after the application of the proposed resource allocation algorithm. Thus, perhaps there exist better ways of modifying the model to adjust the compression rate to make the accuracy better, but this is not the focus of this paper.

Figs. \ref{fig:reconstructed_images} and \ref{fig:rho_k3_acc} are both based on the COCO dataset \cite{coco_dataset}, which is one of the most widely used datasets for object detection in computer vision.
% Fig. \ref{fig:reconstructed_images} illustrates that with the decrease of the compression rate $\rho$, the reconstructed image based on the compressed information gets blurrier and blurrier.
% Figs. \ref{fig:reconstructed_images}(a) and (e) are from the COCO dataset \cite{coco_dataset}. 
Figs. \ref{fig:reconstructed_images}(b)--(d) are reconstructed images based on the compressed information of the original image (a) under different $\rho$. Additionally, Figs. \ref{fig:reconstructed_images}(f)--(h) are reconstructed from the compressed information of (e) based on different $\rho$. The effect of different $\rho$ on the quality of image recovery can also be seen with the naked eye.
As $\rho$ decreases, reconstrcuted figures become blurrier. In Figs. \ref{fig:reconstructed_images}(d) and (h), where $\rho=0.2$, the background appears noticeably noisy, and the contours of objects are less distinct. This degradation aligns with our expectation that lower $\rho$ reduce the transmitted information, leading to blurrier reconstructions.
Overall, these visual results confirm that $\rho$ is a critical parameter: small changes in the compression rate can have a pronounced effect on image clarity and richness. This, in turn, has direct implications for downstream tasks like object detection, as clarity and detail are key drivers of higher accuracy.

Fig. \ref{fig:rho_k3_acc} displays an obvious trend that the model accuracy is influenced by the change of $\kappa_3$ and $\rho$. 
As shown in Fig. \ref{fig:rho_k3_acc}(a), when $\kappa_3$
is small, the optimization may favor faster or more energy‐efficient SemCom transmission at the expense of image quality, leading to lower $\rho$. While when $\kappa_3$ becomes large,
accuracy is prioritized. It results in larger $\rho$ and thus better image quality.
% it is obvious that as $\kappa_3$ increases, the selection of $\rho$ increases since $\kappa_3$ stresses the accuracy and larger $\rho$ implies higher accuracy. 
In Fig. \ref{fig:rho_k3_acc}(b), the accuracy refers to the object detection accuracy. We choose the \mbox{pre-trained} YOLOv3 \cite{redmon2018yolov3} and YOLOv5 \cite{yolov5}, which are \mbox{widely used} models for object detection, for training under different $\rho$.  As $\rho$ becomes larger, the accuracy of the two models both increases. 
It indicates that a higher compression rate produces higher detection precision. This pattern directly links to the visual quality trend in Fig. \ref{fig:reconstructed_images}.
Additionally, as shown in Fig. \ref{fig:rho_k3_acc}(b), we can fit the points of each model in two concave functions. 
These fitting curves also corroborate our hypothesis that the accuracy function is concave, not only for models running on a single device but also applicable to the global model trained through FL. This concavity is not limited to YOLOv3. YOLOv5 also conforms to this pattern.

\subsection{Comparison of Our Proposed Algorithm with the Approximate Exhaustive Search }
Ideally, we would compare the solution of our algorithm with the global optimal solution. 
However, considering that 
the optimization involves variables of various dimensions (e.g., $\boldsymbol{f} \in
\mathbb{R}^{N\times1},\boldsymbol{p}\in
\mathbb{R}^{N\times k}$) and some variables are continuous, the solution searching space is large.
This means exhaustively searching all available domains to find the exact global optimum is impossible. Hence, we propose an approximate exhaustive search method on a toy example to find a solution for a fair comparison.

% use a simple example and  
% the optimization variables have one $N\times 1$, two $N\times K$ variables and a single variable, and some variables are continuous. Hence, the solution space is large. This means exhaustively searching all domains to find the exact global optimum is impossible. However, we try to use a simple example and an approximate exhaustive search method to find a solution for comparison.

Particularly, we set a toy example with $4$ devices and $5$ subcarriers. In our approximate exhaustive search method, the frequency range $f_n$ is from $0.1$ to $2$ GHz,
% , and it is traversed at intervals of $0.005$ GHz. 
and we traverse it in increments of $0.1$ GHz.
The power range $p_{n,k}$ is from $10$ to $20$ dBm, with intervals of $2$ dBm. The compression rate $\rho$ is searched from $0.1$ to $1$ in increments of $0.1$. Thereafter, the time complexity of the approximate exhaustive search is: 
\begin{align}
4 \cdot (\frac{2-0.1}{0.1})^4\cdot (\frac{20-10}{2})^5 \cdot (\frac{1-0.1}{0.1}) \approx 1.5\times10^{10}.
\end{align}
% Reducing the search interval will further increase the time complexity.
% The search results are shown in Table \ref{tab:comparison-exhuastive}.
% , where all weight parameters are set to $1$.

\begin{table}[t!]
\centering
% \rowcolors{1}{myblue}{myblue}  % Apply color to all rows
\caption{The comparison of the proposed algorithm with the approximate exhaustive search.}
\begin{tabular}{|c|c|c|}
\hline
\textbf{Method}    & \begin{tabular}[c]{@{}c@{}}\textbf{Objective function value}\\ \big(i.e., Eq. (\ref{problem:original})\big)\end{tabular} & \textbf{Runtime} (s)
\\ \hline
Equal Allocation& $8.36$ & - \\ \hline
Proposed  & $1.05$ & $2.23$ \\ \hline
\begin{tabular}[c]{@{}c@{}}Approximate \\exhaustive search \end{tabular}    & $0.29$ & $121.37$  \\ \hline

\end{tabular} \label{tab:comparison-exhuastive}
\end{table}
% \begin{table*}[t!]
% \centering
% % \rowcolors{1}{myblue}{myblue}  % Apply color to all rows
% \captionsetup{font={color=blue}}
% \caption{The comparison of the proposed algorithm with approximate exhaustive search.}
% \begin{tabular}{|c|c|c|c|c|c|c|}
% \hline
% \textbf{}        & \textbf{Total energy} & \begin{tabular}[c]{@{}c@{}}\textbf{FL energy}\\($\sum_{n\in\mathcal{N}}E_n^t+E_n^c$)\end{tabular} & \begin{tabular}[c]{@{}c@{}}\textbf{SemCom energy}\\($\sum_{n\in\mathcal{N}}E_n^{sc}$) \end{tabular} & \textbf{FL time} & $\rho$    & \begin{tabular}[c]{@{}c@{}}\textbf{The value of the objective function}\\ \big(i.e., Eq. (\ref{problem:original})\big)\end{tabular} \\ \hline
% Equal Allocation& $0.1076$                & $0.0073$    & $0.1003$        & $1.1223$  & $0.01$   & $-0.2126$  \\ \hline
% Proposed  & $2.8459$   & $0.0855$    & $2.7604$         & $  1.3985$   & $0.0339$ & $0.5587$  \\ \hline
% \begin{tabular}[c]{@{}c@{}}Approximate \\exhaustive search \end{tabular} & $0.1076$                & $0.0073$    & $0.1003$        & $1.1223$  & $0.01$   & $-0.2126$  \\ \hline

% \end{tabular} \label{tab:comparison-exhuastive}
% \end{table*}

% $32.34$

% $5$
Table \ref{tab:comparison-exhuastive} presents the comparison results, and we admit that 
the approximate exhaustive search outperforms our proposed algorithm, but the performance gap is not significant.
Additionally, our proposed algorithm has a definite advantage in terms of solution time and achieves around $54\times$ less runtime. The advantage of our algorithm in terms of runtime will be more pronounced when practical scenarios with higher complexity are considered.

% Although the total energy of the approximate exhaustive search is less, the FL time of our proposed algorithm is shorter. We admit that the approximate exhaustive search may achieve a better result if the search interval is more narrowly divided, but the time complexity will be much higher. 
% \begin{align}
    
% \end{align}

% Hence, from this concise example, it could be observed that our proposed algorithm achieves a better result compared to the approximate exhaustive search since its value of the objective function is worse.

\section{Conclusion} \label{sec:conclusion}
In this paper, we study how to allocate subcarriers, transmission power and CPU frequency of each mobile device in the FedSem system. We formulate an optimization problem that aims to minimize a weighted sum of energy consumption (i.e., the sum of FL energy and SemCom transmission energy), total FL completion time and the model accuracy. To solve the optimization problem, we devise a resource allocation algorithm to allocate the subcarriers, transmission power and CPU frequency for each device. Time complexity and convergence analysis are provided. In addition, in experiments, we investigate our resource allocation algorithm from six aspects: 1) the impact of three weight parameters, 2) the impact of the constraint---the maximum transmission power, 3) the impact of the number of users and subcarriers, 4) the impact of different data samples on our proposed algorithm, 5) the relation between the compression rate and the model accuracy and 6) the comparison of the resource allocation algorithm with the approximate exhaustive search.
Experiments show that our resource allocation algorithm achieves robust performance under different conditions compared to other benchmarks. 

% In addition, the experiments regarding accuracy also demonstrate that an increase in $\kappa_3$ correlates with a corresponding enhancement in accuracy. Furthermore, the accuracy curve can indeed be modeled as a concave function.


\ifCLASSOPTIONcaptionsoff
  \newpage
\fi




\linespread{1.02}

\bibliographystyle{IEEEtran}
\bibliography{ref}

\begin{thebibliography}{10}
\providecommand{\url}[1]{#1}
\csname url@samestyle\endcsname
\providecommand{\newblock}{\relax}
\providecommand{\bibinfo}[2]{#2}
\providecommand{\BIBentrySTDinterwordspacing}{\spaceskip=0pt\relax}
\providecommand{\BIBentryALTinterwordstretchfactor}{4}
\providecommand{\BIBentryALTinterwordspacing}{\spaceskip=\fontdimen2\font plus
\BIBentryALTinterwordstretchfactor\fontdimen3\font minus \fontdimen4\font\relax}
\providecommand{\BIBforeignlanguage}[2]{{%
\expandafter\ifx\csname l@#1\endcsname\relax
\typeout{** WARNING: IEEEtran.bst: No hyphenation pattern has been}%
\typeout{** loaded for the language `#1'. Using the pattern for}%
\typeout{** the default language instead.}%
\else
\language=\csname l@#1\endcsname
\fi
#2}}
\providecommand{\BIBdecl}{\relax}
\BIBdecl

\bibitem{shi2021fromsemcom}
G.~Shi, Y.~Xiao, Y.~Li, and X.~Xie, ``From semantic communication to semantic-aware networking: Model, architecture, and open problems,'' \emph{IEEE Communications Magazine}, vol.~59, no.~8, pp. 44--50, 2021.

\bibitem{luo_sem_2022}
X.~Luo, H.-H. Chen, and Q.~Guo, ``Semantic communications: Overview, open issues, and future research directions,'' \emph{IEEE Wireless Communications}, vol.~29, no.~1, pp. 210--219, 2022.

\bibitem{mcmahan2017communication}
B.~McMahan, E.~Moore, D.~Ramage, S.~Hampson, and B.~A. y~Arcas, ``Communication-efficient learning of deep networks from decentralized data,'' in \emph{Artificial Intelligence and Statistics}.\hskip 1em plus 0.5em minus 0.4em\relax PMLR, 2017, pp. 1273--1282.

\bibitem{wang2022transformer}
Y.~Wang, Z.~Gao, D.~Zheng, S.~Chen, D.~Gunduz, and H.~V. Poor, ``Transformer-empowered 6{G} intelligent networks: From massive {MIMO} processing to semantic communication,'' \emph{IEEE Wireless Communications}, 2022.

\bibitem{pala2023spectral}
S.~Pala, M.~Katwe, K.~Singh, B.~Clerckx, and C.-P. Li, ``Spectral-efficient {RIS}-aided {RSMA} {URLLC}: Toward mobile broadband reliable low latency communication ({mBRLLC}) system,'' \emph{IEEE Transactions on Wireless Communications}, 2023.

\bibitem{xie2021deep}
H.~Xie, Z.~Qin, G.~Y. Li, and B.-H. Juang, ``Deep learning enabled semantic communication systems,'' \emph{IEEE Transactions on Signal Processing}, vol.~69, pp. 2663--2675, 2021.

\bibitem{liu2022extended}
Y.~Liu, S.~Jiang, Y.~Zhang, K.~Cao, L.~Zhou, B.-C. Seet, H.~Zhao, and J.~Wei, ``Extended context-based semantic communication system for text transmission,'' \emph{Digital Communications and Networks}, 2022.

\bibitem{peng2022robust}
X.~Peng, Z.~Qin, D.~Huang, X.~Tao, J.~Lu, G.~Liu, and C.~Pan, ``A robust deep learning enabled semantic communication system for text,'' in \emph{2022 IEEE Global Communications Conference (GLOBECOM)}.\hskip 1em plus 0.5em minus 0.4em\relax IEEE, 2022, pp. 2704--2709.

\bibitem{bourtsoulatze2019deep}
E.~Bourtsoulatze, D.~B. Kurka, and D.~G{\"u}nd{\"u}z, ``Deep joint source-channel coding for wireless image transmission,'' \emph{IEEE Transactions on Cognitive Communications and Networking}, vol.~5, no.~3, pp. 567--579, 2019.

\bibitem{jankowski2020wireless}
M.~Jankowski, D.~G{\"u}nd{\"u}z, and K.~Mikolajczyk, ``Wireless image retrieval at the edge,'' \emph{IEEE Journal on Selected Areas in Communications}, vol.~39, no.~1, pp. 89--100, 2020.

\bibitem{weng2021semantic}
Z.~Weng and Z.~Qin, ``Semantic communication systems for speech transmission,'' \emph{IEEE Journal on Selected Areas in Communications}, vol.~39, no.~8, pp. 2434--2444, 2021.

\bibitem{jiang2022wireless}
P.~Jiang, C.-K. Wen, S.~Jin, and G.~Y. Li, ``Wireless semantic communications for video conferencing,'' \emph{IEEE Journal on Selected Areas in Communications}, vol.~41, no.~1, pp. 230--244, 2022.

\bibitem{vaswani2017attention}
A.~Vaswani, N.~Shazeer, N.~Parmar, J.~Uszkoreit, L.~Jones, A.~N. Gomez, {\L}.~Kaiser, and I.~Polosukhin, ``Attention is all you need,'' \emph{Advances in neural information processing systems}, vol.~30, 2017.

\bibitem{weiss2016survey}
K.~Weiss, T.~M. Khoshgoftaar, and D.~Wang, ``A survey of transfer learning,'' \emph{Journal of Big Data}, vol.~3, no.~1, pp. 1--40, 2016.

\bibitem{yang2023semcomJSAC}
Z.~Yang, M.~Chen, Z.~Zhang, and C.~Huang, ``Energy efficient semantic communication over wireless networks with rate splitting,'' \emph{IEEE Journal on Selected Areas in Communications}, vol.~41, no.~5, pp. 1484--1495, 2023.

\bibitem{li2023resource}
Y.~Li, X.~Zhou, and J.~Zhao, ``Resource allocation for semantic communication under physical-layer security,'' in \emph{2023 IEEE Global Communications Conference (GLOBECOM)}.\hskip 1em plus 0.5em minus 0.4em\relax IEEE, 2023.

\bibitem{liu2023semTCCN}
C.~Liu, C.~Guo, Y.~Yang, and N.~Jiang, ``Adaptable semantic compression and resource allocation for task-oriented communications,'' \emph{IEEE Transactions on Cognitive Communications and Networking}, 2023.

\bibitem{zhou2022joint}
X.~Zhou, J.~Zhao, H.~Han, and C.~Guet, ``Joint optimization of energy consumption and completion time in federated learning,'' in \emph{2022 IEEE 42nd International Conference on Distributed Computing Systems (ICDCS)}.\hskip 1em plus 0.5em minus 0.4em\relax IEEE, 2022, pp. 1005--1017.

\bibitem{dinh2020federated}
C.~T. Dinh, N.~H. Tran, M.~N. Nguyen, C.~S. Hong, W.~Bao, A.~Y. Zomaya, and V.~Gramoli, ``Federated learning over wireless networks: Convergence analysis and resource allocation,'' \emph{IEEE/ACM Transactions on Networking}, vol.~29, no.~1, pp. 398--409, 2020.

\bibitem{zhou2023resource}
X.~Zhou, C.~Liu, and J.~Zhao, ``Resource allocation of federated learning for the {M}etaverse with mobile augmented reality,'' \emph{IEEE Transactions on Wireless Communications}, 2023.

\bibitem{luo2020hfel}
S.~Luo, X.~Chen, Q.~Wu, Z.~Zhou, and S.~Yu, ``{HFEL}: Joint edge association and resource allocation for cost-efficient hierarchical federated edge learning,'' \emph{IEEE Transactions on Wireless Communications}, vol.~19, no.~10, pp. 6535--6548, 2020.

\bibitem{nguyen2020efficient}
V.-D. Nguyen, S.~K. Sharma, T.~X. Vu, S.~Chatzinotas, and B.~Ottersten, ``Efficient federated learning algorithm for resource allocation in wireless {IoT} networks,'' \emph{IEEE Internet of Things Journal}, vol.~8, no.~5, pp. 3394--3409, 2020.

\bibitem{liu2022joint}
S.~Liu, G.~Yu, X.~Chen, and M.~Bennis, ``Joint user association and resource allocation for wireless hierarchical federated learning with {IID} and non-{IID} data,'' \emph{IEEE Transactions on Wireless Communications}, vol.~21, no.~10, pp. 7852--7866, 2022.

\bibitem{chen2020joint}
M.~Chen, Z.~Yang, W.~Saad, C.~Yin, H.~V. Poor, and S.~Cui, ``A joint learning and communications framework for federated learning over wireless networks,'' \emph{IEEE transactions on wireless communications}, vol.~20, no.~1, pp. 269--283, 2020.

\bibitem{yang2020energy}
Z.~Yang, M.~Chen, W.~Saad, C.~S. Hong, and M.~Shikh-Bahaei, ``Energy efficient federated learning over wireless communication networks,'' \emph{IEEE Transactions on Wireless Communications}, vol.~20, no.~3, pp. 1935--1949, 2020.

\bibitem{tong2021federated}
H.~Tong, Z.~Yang, S.~Wang, Y.~Hu, W.~Saad, and C.~Yin, ``Federated learning based audio semantic communication over wireless networks,'' in \emph{2021 IEEE Global Communications Conference (GLOBECOM)}.\hskip 1em plus 0.5em minus 0.4em\relax IEEE, 2021, pp. 1--6.

\bibitem{tongSemComJournal2021}
H.~Tong, Z.~Yang, S.~Wang, Y.~Hu, O.~Semiari, W.~Saad, and C.~Yin, ``Federated learning for audio semantic communication,'' \emph{Frontiers in Communications and Networks}, vol.~2, 2021.

\bibitem{loc2023federated}
X.~N. Loc, Q.~L. Huy, L.~T. Ye, S.~A. Pyae, K.~T. Yan, H.~Zhu, and S.~H. Choong, ``An efficient federated learning framework for training semantic communication system,'' \emph{IEEE Transactions on Vehicular Technology}, 2024.

\bibitem{weiFedSem2023Letters}
H.~Wei, W.~Ni, W.~Xu, F.~Wang, D.~Niyato, and P.~Zhang, ``Federated semantic learning driven by information bottleneck for task-oriented communications,'' \emph{IEEE Communications Letters}, vol.~27, no.~10, pp. 2652--2656, 2023.

\bibitem{li2023CATFL}
G.~Li, Y.~Zhao, and Y.~Li, ``{CATFL}: Certificateless authentication-based trustworthy federated learning for {6G} semantic communications,'' in \emph{2023 IEEE Wireless Communications and Networking Conference (WCNC)}, 2023, pp. 1--6.

\bibitem{xie2023asynchronous}
R.~Xie, C.~Li, X.~Zhou, and Z.~Dong, ``Asynchronous federated learning for real-time multiple license plate recognition through semantic communication,'' in \emph{2023 IEEE International Conference on Acoustics, Speech and Signal Processing (ICASSP)}.\hskip 1em plus 0.5em minus 0.4em\relax IEEE, 2023, pp. 1--5.

\bibitem{xing2023deepSemCom}
H.~Xing, H.~Zhang, X.~Wang, L.~Xu, Z.~Xiao, B.~Zhao, S.~Luo, L.~Feng, and Y.~Dai, ``A multi-user deep semantic communication system based on federated learning with dynamic model aggregation,'' in \emph{2023 IEEE International Conference on Communications Workshops (ICC Workshops)}, 2023, pp. 1612--1616.

\bibitem{xing2023fedDistillSC}
H.~Xing, H.~Ma, Z.~Xiao, B.~Zhao, J.~Peng, S.~Luo, L.~Feng, and L.~Xu, ``{FedDistillSC}: A distributed semantic communication system based on federated distillation,'' in \emph{2023 IEEE 23rd International Conference on Communication Technology (ICCT)}, 2023, pp. 506--510.

\bibitem{liu2023SemCom2023VTC}
J.~Liu, Y.~Lu, H.~Wu, and Y.~Dai, ``Efficient resource allocation and semantic extraction for federated learning empowered vehicular semantic communication,'' in \emph{2023 IEEE 98th Vehicular Technology Conference (VTC2023-Fall)}, 2023, pp. 1--5.

\bibitem{song2023robustFedSem}
Y.~Song, J.~Wang, and Y.~Liu, ``Robust federated learning for image semantic transmission under {Byzantine} attacks,'' in \emph{2023 IEEE/CIC International Conference on Communications in China (ICCC Workshops)}, 2023, pp. 1--6.

\bibitem{10251890}
J.~Chen, J.~Wang, C.~Jiang, Y.~Ren, and L.~Hanzo, ``Trustworthy semantic communications for the metaverse relying on federated learning,'' \emph{IEEE Wireless Communications}, vol.~30, no.~4, pp. 18--25, 2023.

\bibitem{10422907}
J.~Xu, H.~Yao, R.~Zhang, T.~Mai, S.~Huang, and S.~Guo, ``Federated learning powered semantic communication for uav swarm cooperation,'' \emph{IEEE Wireless Communications}, vol.~31, no.~4, pp. 140--146, 2024.

\bibitem{redmon2018yolov3}
J.~Redmon and A.~Farhadi, ``{YOLO}v3: An incremental improvement,'' \emph{arXiv preprint arXiv:1804.02767}, 2018.

\bibitem{coco_dataset}
T.-Y. Lin, M.~Maire, S.~Belongie, J.~Hays, P.~Perona, D.~Ramanan, P.~Doll{\'a}r, and C.~L. Zitnick, ``Microsoft {COCO}: Common objects in context,'' in \emph{European Conference on Computer Vision (ECCV)}, 2014, pp. 740--755.

\bibitem{cambinigeneralized}
A.~Cambini and L.~Martein, \emph{Generalized convexity and optimization: Theory and applications}.\hskip 1em plus 0.5em minus 0.4em\relax Springer Science \& Business Media, 2008, vol. 616.

\bibitem{jong2012efficient}
Y.~Jong, ``An efficient global optimization algorithm for nonlinear sum-of-ratios problem,'' \emph{Optimization Online}, pp. 1--21, 2012.

\bibitem{nam2019ofdma}
N.-T. Le, L.-N. Tran, Q.-D. Vu, and D.~Jayalath, ``Energy-efficient resource allocation for {OFDMA} heterogeneous networks,'' \emph{IEEE Transactions on Communications}, vol.~67, no.~10, pp. 7043--7057, 2019.

\bibitem{zhao2023human}
J.~Zhao, L.~Qian, and W.~Yu, ``Human-centric resource allocation in the {Metaverse} over wireless communications,'' \emph{IEEE Journal on Selected Areas in Communications}, 2023.

\bibitem{yolov5}
\BIBentryALTinterwordspacing
G.~Jocher, A.~Chaurasia, A.~Stoken, J.~Borovec, NanoCode012, Y.~Kwon, TaoXie, K.~Michael, J.~Fang, imyhxy, Lorna, C.~Wong, Z.~Yifu, A.~V, D.~Montes, Z.~Wang, C.~Fati, J.~Nadar, Laughing, UnglvKitDe, tkianai, yxNONG, P.~Skalski, A.~Hogan, M.~Strobel, M.~Jain, L.~Mammana, and xylieong, ``ultralytics/yolov5: v6.2 - {YOLOv5} classification models, {A}pple {M1}, reproducibility, clear{ML} and deci.ai integrations,'' Aug. 2022. [Online]. Available: \url{https://doi.org/10.5281/zenodo.7002879}
\BIBentrySTDinterwordspacing

\end{thebibliography}


%\vfill


\begin{appendices}

\end{appendices}

\end{document}